\documentclass[a4paper,12pt]{article}%
\usepackage{amsmath}%
\usepackage{txfonts}%
\usepackage{amsfonts}%
\usepackage{amssymb}%
\usepackage{graphicx}
\usepackage{wrapfig}
\usepackage{color}
\usepackage{setspace}
\usepackage{esint}
\usepackage[rightcaption]{sidecap}
\usepackage{wrapfig}
\usepackage{lipsum}
\usepackage{caption}
\usepackage{longtable}
\usepackage[T2A,T1]{fontenc}
\usepackage[utf8]{inputenc}
\usepackage[russian,english]{babel}

\usepackage{etoolbox}
\makeatletter
\patchcmd{\@maketitle}{\begin{center}}{\begin{flushleft}}{}{}
\patchcmd{\@maketitle}{\begin{tabular}[t]{c}}{\begin{tabular}[t]{@{}l}}{}{}
\patchcmd{\@maketitle}{\end{center}}{\end{flushleft}}{}{}
\makeatother

\numberwithin{equation}{section}

\newcommand{\rr}{\color{red}}
\newcommand{\be}{\begin{equation}}
\newcommand{\ee}{\end{equation}}

\usepackage{anysize}
\marginsize{2cm}{2cm}{2cm}{2cm}
\usepackage{mathptmx}
-1in\relax 

\begin{document}

\begin{center}
\LARGE
\textbf{Feasibility Study For Multiply Reusable\\ Space Launch System.}
\normalsize
\vskip1cm
\begin{tabular}{l}
\hskip1cm
\Large Mikhail V. Shubov         \\
\normalsize \\
\hskip1cm University of MA Lowell    \\
\hskip1cm One University Ave,        \\
\hskip1cm Lowell, MA 01854           \\
\hskip1cm E-mail: mvs5763@yahoo.com  \\
\end{tabular}
\end{center}

\tableofcontents

\begin{center}
  \textbf{Abstract}
\end{center}
\begin{quote}
A novel concept of orbital launch system in which all stages are reusable is presented.  The first two stages called Midpoint Delivery System (MPDS) deliver the next stages to a midpoint.  A midpoint is defined by an altitude of 100 $km$ to 120 $km$ and horizontal velocity of 2.8 $km/s$ to 3.2 $km/s$.  MPDS stages decelerate in the atmosphere and perform vertical landing on barges.  These stages can be reused daily for many years.  The payload is delivered from the midpoint to a 400 $km$ Low Earth Orbit by one or two stage rocket called Midpoint to Orbit Delivery System (MPTO).  All of MPTO engines are delivered to LEO.  These engines do not return to Earth themselves.  They are returned to Earth in packs of 50 to 100 by a Reentry Vehicle.  Overall, the fully and multiply reusable launch system should deliver payload to LEO for \$300 to \$400 per $kg$\\
\textbf{Keywords:}
\end{quote}

\section{Introduction}

Space Colonization holds the promise of opening almost unlimited resources and ushering the New Era for Human Civilization.  The resources contained within the Solar System can support a civilization much more populous and advanced than the modern one.  The most important resource for modern industry and civilization is energy \cite{kard, E1}.  The solar energy in space is almost limitless -- the Sun's thermal power is
$3.86 \cdot 10^{26}\ W$ \cite[p.14-2]{crc}.  A future civilization, which would harvest 10\% of that power with 10\% efficiency, will have energy production of
$3.4 \cdot 10^{25}\ kWh/year$.  The scheme for harvesting this energy is called the Dyson Sphere \cite{Dyson}.  The Dyson Sphere would consist of a multitude of solar energy harvesting space stations orbiting the Sun.  The current global energy production in all forms is $7 \cdot 10^{13}\ kWh/year$ \cite{wes} -- 500 billion times less.  The Asteroid Belt contains almost unlimited material resources -- about $3 \cdot 10^{18}\ tons$ of material composed of metal silicates, carbon compounds, water, and pure metals \cite{ABM}.  Most asteroids are of a carbonaceous type \cite{Asteroids}.  Carbon is very useful for production of food for space travelers, production of fuel for propulsion within space, and for production of plastics for space habitat structures.  High quality steel is also an abundant resource in space, e.g., asteroid 16 Psyche contains $10^{16}\ tons$ of nickel-rich steel \cite{Psycho}.  It has been calculated that Solar System resources can easily sustain a population a million times greater than the global population of today \cite{skymine}.  Thus, colonization of the Solar System will be an extraordinary important step for Humankind.

The colonization of the Solar System was originally proposed by Konstantin Tsiolkovsky in 1903 \cite{Tsialkovski}.  By the 1970s, elaborate road maps for space colonization were published \cite{habitats,1974}.  In the previous work, I have demonstrated the feasibility of a Gas Core Reactor (GCR) acting as a power source for a deep space transportation system \cite{Shubov2}, which would be very useful for space colonization.

No major step in development of Humankind has been easy so far, and the same will likely hold true for coming advances.  The colonization of the Solar System may present Humankind with many known and yet unknown challenges which must be overcome.  The first challenge is reducing the cost of the first step -- transfer of payload and astronauts from Earth to Low Earth Orbit (LEO).  This work will focus on overcoming this challenge.  According to Ancient Chinese wisdom, even the longest journey begins with the first step.

In this work, we demonstrate the feasibility of a launch system which would accomplish the first step of Space Colonization -- delivery of astronauts and payload to Lower Earth Orbit at reasonable cost.
ODS consists of a Midpoint Delivery System (MPDS) and a Midpoint to Orbit Delivery System (MPTO).  MPDS delivers MPTO and a payload to a \textbf{midpoint} in an orbital ascent.  A midpoint is a combination of an altitude of about 100 $km$ to 120 $km$ and an initial velocity of 2.8 $km/s$ to 3.3 $km/s$.  MPTO delivers the payload to an orbital station at 400 $km$ orbit.

MPDS consists of two stages, both of which make a vertical landing.
MPDS reusable every day and have a service life of at least 4,000 launches.
MPTO stages have reusable fuel tanks.
The engine and any costly part of MPTO is returned to Earth via a Reentry Vehicle.
These parts should be reusable.

The liftoff mass of the MPDS-MPTO space delivery system is 2,000 $tons$. The MPDS-MPTO system delivers 5.1 $tons$ payload in a 600 $kg$ container to a 400 $km$ LEO.  The giant engines of the MPDS stages use liquid methane -- liquid oxygen combination, which costs an average of \$184 per ton.  Moderately sized engines of MPDS use hypergolic propellant, which costs about \$2,000 per ton.  Orbital delivery cost should be about \$300 to \$400 per $kg$ payload.

The MPDS-MPTO system should bring the Humankind closer to the era of Space Colonization and unlimited resources.  It should be a great step for Humankind.  The three military systems seem to be destructive in their very purpose.  Nevertheless, it should bring more knowledge to Rocket Science at lower cost.  Thus, all four systems will serve Humankind.

In Section 2, we describe the state of the art and proposed orbital delivery systems.  We trace the evolution of launch costs from 1957 to 2018.  In 1957, the cost of placing a payload into Low Earth Orbit (LEO) was about \$900,000 per $kg$ in 2018 dollars.  By 1965, the cost of placing payload into orbit has fallen to a range of \$5,000 to \$20,000 per $kg$.  Costs stayed in that range until SpaceX became able to deliver payload at \$1,400 per $kg$.  In Subsection 2.3 we discuss the Single Stage to orbit concept and demonstrate its impossibility.  In Section 2.4, we present our concept of MPDS-MPTO system.

In Section 3, we present the physics of rocket flight and propulsion.  In Subsection 3.1, we describe the general physics of rocket motion.  In Subsection 3.2, we describe physics of liquid rocket engines.  In Subsection 3.3, we describe rocket propellants and oxidizers.  In Subsection 3.4, we present the costs of different propellants, as well as the total fuel bill for space launch.

In Section 4, we describe performance of the launch system.
In Subsection 4.1, performance of MPDS is described.
In Subsection 4.2, performance of MPTO is described.

In Section 5, we evaluate the rate of aerodynamic heating of MPDS stages during descent.
In Subsection 5.1, we evaluate the rate of laminar heating.
In Subsection 5.2, we evaluate the rate of turbulent heating.
In Subsection 5.3, we describe heat sink shields.
In Subsection 5.4, we evaluate heating and deceleration rates of MPDS stages.
In Subsection 5.5, we describe the heat shield of the Space Shuttle and explain its failure.

\section{State of Art and Proposed Orbital Delivery Systems}

\subsection{Historical Launch Costs}
\hspace{.5cm} In the period from 1957 up to 2015, there have been 5,510 space launches, of which 5,046 were successful.  The decade with most (1,231) space launches was the 1970s  \cite{SLR}.  During the years 1990 -- 2010, 16,200 tons of payload have been launched into space and 65\% of that mass returned to the Earth with the Space Shuttle \cite{TotM}.

One of the main parameters for a launch vehicle is the cost of placing a payload into orbit.  The cost is proportional to payload mass, and it is measured in dollars per $kg$.  The evolution of launch cost is illustrated in \cite{LCost}:
\begin{center}
\includegraphics[width=12cm]{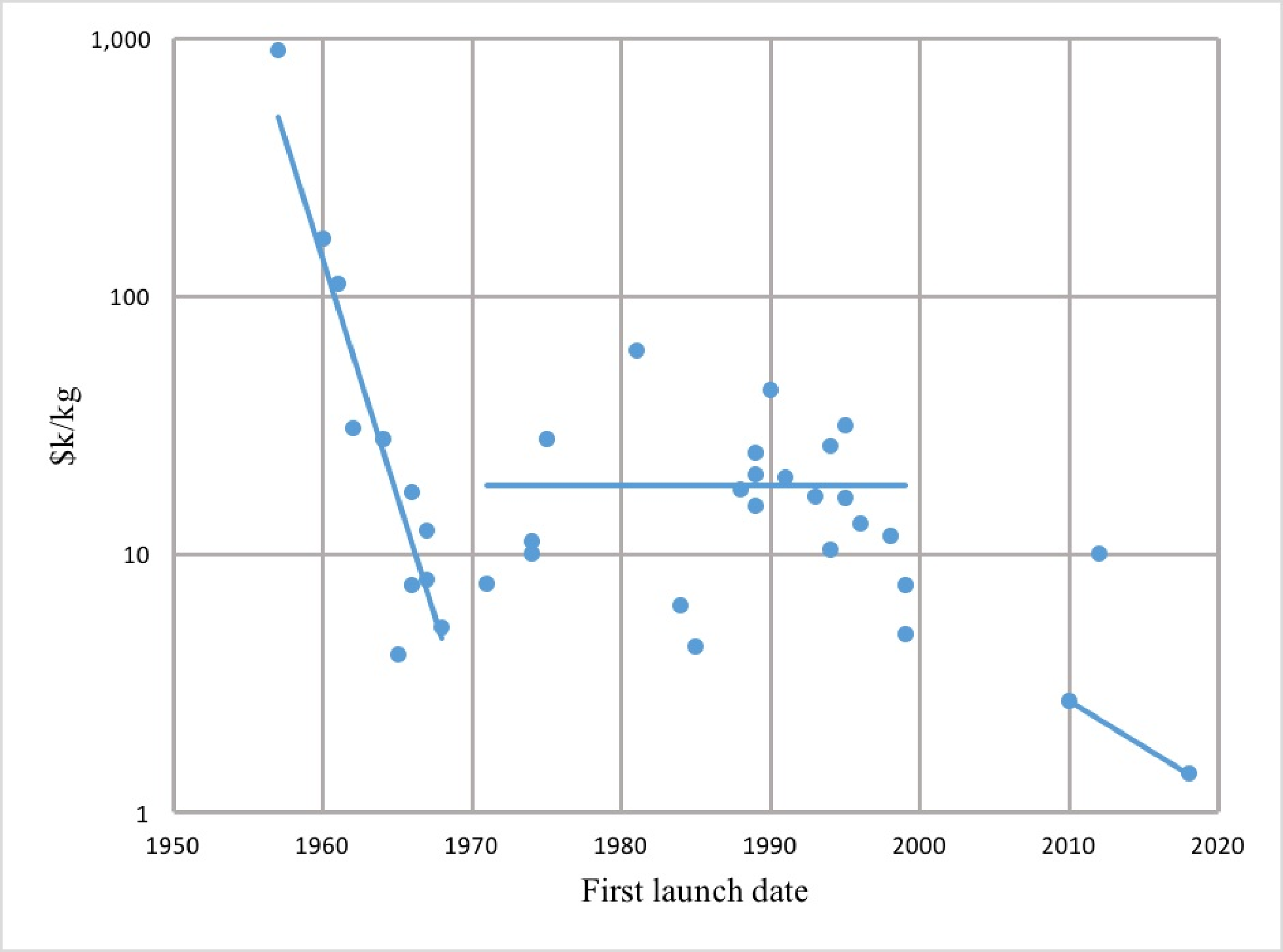}
\captionof{figure}{Launch Cost Evolution \label{F01}}
\end{center}

The cost of placing payload into the Low Earth Orbit (LEO) by the first rockets was very high.  For Vanguard launched in 1957 it was \$895,000 per $kg$ in 2018 dollars.  At first the cost dropped rapidly -- Delta E launched in 1960 had a cost of \$168,000 per $kg$ in 2018 dollars \cite[p.8]{LCost}.  Saturn V Rocket used in the Moon expedition in 1969 cost which is \$728 million in 2018 dollars \cite[p.8]{LCost}.  It could have placed 140 tons into LEO \cite[p.8]{LCost}.  Thus, by 1969, very heavy launch vehicles could place a payload into an orbit at \$5,200 per $kg$ in 2015 dollars.

In the early 1970s, most experts believed that in a few years, the cost of placing a payload into an orbit would have been \$2,200 per $kg$ in 2015 dollars \cite[p.35]{Shuttle}.  By 1980s, space launch should have cost \$400 -- \$600 per $kg$ payload in 2015 dollars \cite[p.viii]{guide}.  Unfortunately, the progress was stalled, and launch costs remained at an average of \$18,500 per $kg$ up to about 2010 \cite[p.8]{LCost}.  In 2002, the cost of LEO delivery has been \$6,900 per $kg$ for Delta4 Heavy rocket and \$18,400 per $kg$ for Titan 4B rocket \cite[p.211]{Cost1}.

In 2016, Ariane 5 ECA rocket delivered payload to LEO for \$8,500 per $kg$ \cite[p.74]{Raport01}.  Even in 2019, old and expensive launch systems are used along with the advanced SpaceX launch systems discussed in the next subsection.  It takes time for new technology to replace the old one.

\subsection{Launch Cost Reduction}
The combination of high anticipations in early 1970s and very limited progress by mid 2000s was very disappointing.  It could have taken a very long time for any progress to happen.

A giant breakthrough was accomplishes by SpaceX company.  Already by 2009, their Falcon 9 rocket delivered payload to LEO for \$2,700 per $kg$.  The next great step was the introduction of the reusable first stage.  On December 21, 2015, SpaceX made a huge step in History when the first stage of Falcon 9 spacecraft returned to the launching pad \cite[p.1]{SLVD}.  During 2016, SpaceX has successfully landed six first stage boosters \cite{Falcon2}.  By July 2019, there have been 34 successful first stage returns out of 40 attempts \cite{Falcon3}.  By 2018, SpaceX was offering LEO delivery at \$1,400 per $kg$ via Falcon Heavy \cite[p.8]{LCost}.  A two stage spacecraft with fully reusable stages will deliver payload to LEO for \$1,000 per $kg$ \cite[p.33]{Falcon1}.

Even though much has been achieved, more remains to be done.  According to the interview with SpaceX CEO Elon Musk \cite{Falcon3},

\begin{quote}
by late 2014, SpaceX determined that the mass needed for a re-entry heat shield, landing engines, and other equipment to support recovery of the second stage was at that time prohibitive, and indefinitely suspended their second-stage reusability plans for the Falcon line.
\end{quote}

\subsection{Single Stage to Orbit (SSTO) Concept }

A Single Stage to Orbit (SSTO) vehicle is supposed to be a single-stage rocket which would deliver payload to Low Earth Orbit.  It must also be able to return to Earth from the orbit.  No SSTO has been built so far.  The very concept seems to be impossible with the state of the art technology.  The main problem with SSTO is the fact that all components of the system would have to be stretched to the limit.  Extremely high loads on all components of the system would make it extremely expensive and unreliable.

Space Shuttle came closest to SSTO \cite{Shuttle1}.  It can be called a 1.5 stage to orbit vehicle.  The Shuttle has flown 135 times for a total of \$209 Billion in 2010 dollars \cite{Shuttle2}, which is \$1.54 Billion per flight.  That is 20 times more expensive than the original plan for program cost.  The Shuttle weighs 78 $tons$, and it can deliver 27.5 $tons$ to LEO \cite{Shuttle1}.  Given the high cost per ton for LEO delivery, the Shuttle can be regarded as a massive failure.

The Shuttle Program also put a stop to the Saturn V program.  Saturn V rocket was last launched in 1973, even though the program was fully developed, and two unused rockets remained \cite{Saturn1}.  Saturn V could deliver 140 tons to LEO for \$1.16 Billion per launch in 2016 dollars \cite{Saturn1}.  Saturn rocket took two stages to get to LEO.  Even if a space capsule took a significant part of the payload, the system was superior to the Space Shuttle.  The main reason for the superiority of the earlier Saturn system over the later Shuttle system is the fact that the parameters of a two stage vehicle do not have to be stretched to the limit like those of a 1.5 stage one.  Low safety margins of rockets of high performance significantly increase their design cost \cite[p. 17]{SMargin}.

\subsection{Non-Rocket Space Launch}

There are several concepts for non-rocket space launch.  A \textbf{skyhook} is a long tether extended from an orbital vehicle.  It grabs a payload on a suborbital flight and delivers it to an orbit \cite{SkyHook}.  In some designs, the skyhook lifts the payload from a hypersonic airplane \cite{HASTOL}.

A \textbf{launch loop} is a 2000 $km$ long runway built 80 $km$ above Earth's surface \cite{LaunchLoop}.  This loop is supported by \textbf{inflatable space towers}.   These are towers made out of very strong impermeable fiber and filled with helium gas \cite[p.72-92]{NRSL1}.  Towers tens of kilometers tall can also be built out of pressurised blocks.  These blocks would be composed of very strong fibers and pressurized by helium gas \cite[p.12-13]{SpaceElevator}.  A launch loop can deliver about 5 million tons of payload per year to LEO \cite{LaunchLoop}.

A \textbf{space gun} accelerates a payload by electromagnetic force in an inclined tunnel and launches it into space \cite{SpaceCannon}.  One form of a space cannon consists of a tube 4 $m$ in diameter and several $km$ long.  The tube is filled with an explosive gas combination -- like oxygen and methane.  The transport enters the tube at a velocity of 1.5 $km/s$.  The transport has scramjet  engines.  These engines accelerate the transport by burning the gas mixture.  The space gun can deliver 50,000 tons payload per year at a cost of \$2.50 per $kg$.  Neither astronauts nor fragile electronics can be delivered to space by a gun due to high launch acceleration \cite[p.104-121]{NRSL1}.

A \textbf{star tram} is an upgraded version of a space gun.  This space gun consists of a vacuum tube about 130 $km$ long.  The firing end of the tube is located either at a mountain top about 6 $km$ tall or is suspended in atmosphere at 10 $km$ by lifting gas balloons.  The end of the tube is inclined at 10$^o$ to horizontal.

The transport is a streamline body 2 $m$ in diameter and 13 $m$ in length.  Loaded transport has a mass of 40 $tons$.  The transport is accelerated within a vacuum tunnel by a linear induction motor.  It levitates due to magnetic field and never touches the surface.  The transport leaves the electromagnetic cannon at 8.9 $km/s$.  It cruises out of the atmosphere and uses a small rocket engine to get into Earth's orbit.

A star tram system would cost about \$100 Billion.  It would launch about 150,000 tons of payload per year into orbit.  The transports experience 30 $g$ acceleration, thus they can not be used to transport passengers or fragile cargo.  Nevertheless, they can transport both sturdy cargo and fuel \cite{StarTram2010,StarTram2010A}.

A \textbf{space elevator} consists of a long cable.  The bottom is attached to some point on Earth's Equator.  The top is attached to a counterweight above geosynchronous orbit.  The elevator rotates along with the Earth.  The centripetal force experienced by the counterweight prevents it from falling.  Space Elevator would be able to deliver payload to deep space at \$1.30 per $kg$ \cite[p.25]{NRSL1}.  Unfortunately, no material capable of withstanding the cable tension exists \cite{SpaceEl1,SpaceEl2}.

An \textbf{orbital ring} is a structure of an artificial ring placed around Earth or any other planet.  The ring is generally, but not necessarily, placed around a planet's equator.  It is circular.  The ring has the same center as the planet itself.  The ring consists of two parts.
The \textbf{stationary tube} is a tube extending for the full length of the ring.  If the ring is above the planet's equator, the stationary tube is stationary with respect to the equatorial planet surface.
The \textbf{moving cable} is inside the stationary tube.  This cable rotates around the planet at a superorbital velocity.  Centripetal force provided by the rotation of the moving cable supports the stationary tube and prevents it from falling.
The moving cable is not touching the stationary tube.  It is supported by a magnetic cushion \cite{OrbitalRing}.  Similar cushions are used in magnetic levitation (MagLev) trains \cite{MagLev01,MagLev05}.  The orbital ring has LEO delivery capacity of 40 billion tons per year \cite{OrbitalRing}.

None of the none-rocket space launch projects has been attempted so far.
None of them are even close to the experimental stage of development \cite{NRSL1}.
In the author's opinion, some aforementioned methods will be useful during the later stages of Solar System Colonization.
First, a star tram transporting about 150,000 tons per year to LEO would be built.
Second, a launch loop delivering about 5 million $tons$ per year to LEO would be constructed.
Third, an orbital ring delivering about 40 billion tons of payload per year to LEO would be built.
At the initial stages of Solar System Colonization, reusable rockets must be used.  The MPDS-MPTO system is described in the following sections.

\section{Multiply Reusable Launch System -- Purpose and Concept}
\subsection{Early Stages of Solar System Colonization}

Current Space Industry brings all kinds of communication, observation and GPS satellites to orbits of destination.  It provides the lifting power for deep space probes \cite{Raport01}.  It also transports astronauts to space.  Since the 1960s, 574 astronauts have been to space.  In February 2020, there are 6 astronauts in space \cite{Astronauts}.

Current Space Industry is insufficient for Space Colonization.  First, it has relatively low gross payload capacity.  Between 2011 and 2018, all Nations launched about 380 $tons$ payload per year into space -- 210 $tons$ into LEO, 130 $tons$ to GEO, and 40 $tons$ to other orbits \cite{LaunchStat1}.  This payload is equivalent to delivery of 670 $tons$ per year to LEO \cite{Raport01}.  Second, it has no uniform standard.  In the 2010s there have been 886 space launches with 45 failures.  There are tens of rocket types all over the World \cite{SpaceLaunchSum}.  Third, orbital delivery is still very expensive.  Most payload is delivered by launchers with price tags of \$10,000 to \$20,000 per $kg$ placed in LEO \cite{Raport01}.

Solar System Colonization is likely to start with the Moon.  The Moon is close to Earth, and it has vast resources.  Lunar poles contain at least 430 million tons of ice \cite[p.326]{LunarIce}.
Other sources estimate the ice deposits at lunar poles as 660 million tons \cite[p.368]{LunaResources1}.  In most polar craters which contain ice, water makes up 2.5\% of the top 1.5 meters of regolith \cite[p.329]{LunarIce}.
Even though ice is plentiful on Earth, in space it is an important resource.  It can be used to manufacture propellant.  Even though, propellant can be inexpensive on Earth, in space it has much higher value due to delivery cost.

Lunar surface contains 6\% to 10\% iron oxide and 0.09\% to 0.18\% chromium oxide \cite[p. 400]{LunarAu}.  These metals are of primary importance for space manufacturing.  Lunar surface has rich deposits of gold and platinum-group metals.  It contains about 3.0 $ppm$ (parts per million by weight) gold, and 5 $ppm$ to 8 $ppm$ iridium \cite[p. 400]{LunarAu}.

In order to start the colonization on the Moon, we have to bring an initial payload of about 1,000 tons and 25 astronauts.  Moreover, about 10 astronauts would have to be initially present on the site \cite[p.207]{LunarBase1}.  To sustain a lunar base, about 650 $tons$ of payload would have to be brought from Earth to LEO each year \cite[p.132]{LSBchapter03}.

Delivery of payload to the Moon would involve delivery of considerably greater mass of payload to LEO.  The $\triangle v$ for a trip from LEO to Moon with soft landing is 6.1 $km/s$.  The $\triangle v$ for a similar lunar trip and return to Earth is 8.6 $km/s$.  It takes 5 $kg$ to 10 $kg$ payload delivered to LEO in order to deliver 1 $kg$ payload to Moon with soft landing.  It takes 25 $kg$ to 100 $kg$ LEO payload for a similar lunar trip with return to Earth \cite[p.130]{RPE}.

Overall, setting up an industrial base on the Moon would initially require annual delivery of 650 $tons$ payload to the Moon.  This in turn would require annual delivery of 3,300 $tons$ to 6,500 $tons$ payload to LEO.  Later stages of Moon colonization as well as initial stages of colonizing asteroids and other planets would have even greater LEO launch requirements.

In order for LEO payload requirements to be met, a multiply reusable launch system must be developed.  This system must be able to deliver a large mass of payload to LEO at a relatively low cost.  The MPDS-MPTO system we are proposing should be able to deliver 300 payloads of 5.1 $tons$ each per year.  The cost should be \$300 to \$400 per $kg$ payload.  The system is described below.

\subsection{The Concept of Multiply Reusable Launch System Consisting of MPDS and MPTO}
\begin{wrapfigure}{R}{3cm}
\includegraphics[width=3cm,height=8cm]{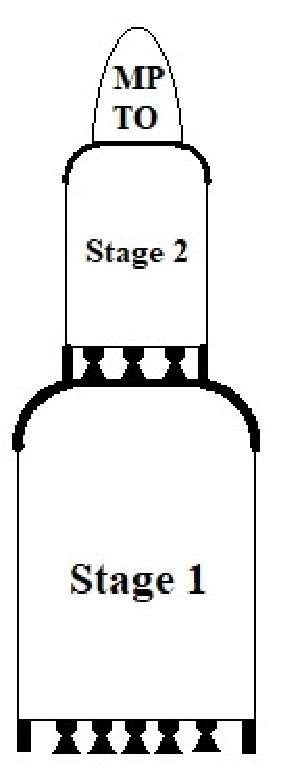}
\captionof{figure}{Launch System Drawn to Scale \label{F02}}
\end{wrapfigure}
The launch system we are suggesting consists of two parts.  Midpoint Delivery System (MPDS) delivers the Midpoint to Orbit Delivery System (MPTO) and the payload to a midpoint in orbital ascent.  MPTO delivers the payload from a midpoint to orbit.  A midpoint is a combination of an altitude of about 100 $km$ to 120 $km$ and an initial velocity of 2.8 $km/s$ to 3.3 $km/s$.  In this work, a midpoint is not strictly defined.  Most likely, first versions of MPDS will deliver MPTO and payload to a ``low" midpoint, while later versions will deliver it to a ``high" midpoint.

MPDS consists of two multiply reusable stages.  The most important feature and advantage of MPDS stages is their durability and wide safety margins, enabling multiple reusability.  Reusability is achieved by their ability to land.  Multiple reusability is achieved by a significant reduction of MPDS stages' workload relative to other launch systems.  As we explain in Subsection 3.2, the main factor defining the workload of a rocket stage is $\triangle v$.  State of the art two-stage rockets have $\triangle v \approx 4,700\ m/s$.  In MPDS, Stage 1 has $\triangle v = 2,600\ m/s$  and Stage 2 has $\triangle v = 1,850\ m/s$ .

Low $\triangle v$ requirement enhances durability and safety margins of both the rocket body and the engines.  The rocket body can have higher \textbf{empty mass ratio}, which is the mass ratio of empty to fuelled rocket.  Stage 1 of MPDS has an empty mass ratio of 14.4\%.  Stage 2 of MPDS has an empty mass ratio of 23.4\%.  Currently, used rockets listed in Appendix B have empty mass ratios of 6.3\% to 9.6\%.  Heavier rockets have lower empty mass ratios, thus both Stage 1 and Stage 2 have empty mass ratios at least twice as high as the other rockets in their corresponding mass classes.  High empty mass ratios allow for high safety margins.

Low $\triangle v$ requirement allows rocket engines to have lower exhaust velocity.  Lower exhaust velocity requirement allows the rocket to have lower combustion temperature.  That is achieved by burning a fuel-rich combination of fuel and oxidizer.  MPDS engines use liquid methane fuel and liquid oxygen oxidizer.  These engines use oxidizer to fuel mass ratio of 1.89 to 1.  This combination produces a combustion chamber temperature of 2,100 $^o$C.

Several other rockets proposed by major companies also use liquid methane fuel and liquid oxygen oxidizer. Prometheus would use an oxidizer to fuel mass ratio of 3.5 to 1 \cite{Prometey}.  Raptor designed by SpaceX uses a similar ratio \cite{Raptor}.  This enables them to raise the exhaust velocity by 14\% to 15\% relative to our rocket.  At the same time, the combustion chamber temperature is raised to 3,330 $^o$C.  Extremely hot combustion damages the combustion chamber.

MPDS stages experience relatively low heating upon reentry.  Stages 1 and 2 use metal heat sink shields.  Metal heat sink shields can tolerate much lower thermal loads than either ablative heat shields or ceramic heat shields used on the Shuttle.  The main advantage of metal heat sink shields is their durability and multi-reusability.

Material requirements for heat shields are presented in Subsection 4.3.  Stage 1 heat shield has a mass of 6 $tons$, while Stage 2 heat shield has a mass of 5 $tons$.  As we mention in Subsection 4.1, the total inert mass of Stage 1 is 230 $tons$ and the inert mass of Stage 2 is 75 $tons$.  Thus, the heat shields would constitute small fractions of stage masses.

The Midpoint to Orbit Delivery System would consist either of two pressure-fed hypergolic rocket stages or one pump-fed hypergolic rocket stage.  In either case, the system should be relatively inexpensive.

MPTO engine would not be able to return from the orbit on its own.  That does not make it non-reusable.  MPTO engines are accumulated on a space station.  Once about 50 to 100 MPTO engines arrive, they are packed and loaded onto a Reentry Vehicle.  This vehicle takes them back to Earth for further reuses.

\section{Rocket Motion and Propulsion}

\subsection{Rocket Flight Physics}

The physics of rocket flight for rocket design is extremely complicated.  It has to take into consideration many predictable and some unpredictable forces acting on a rocket.  It has to make a detailed calculation of air resistance and engine thrust profiles over time.  Designing a rocket involves a lot of theoretical and experimental work -- at least several tens of thousands of expert-years \cite[p.116]{SLVD}.  Producing a feasibility study requires much less detailed analysis and much more simplified equations of rocket motion.

\subsubsection{Rocket Motion}

Prior to discussing the equations of rocket motion, we introduce the system of coordinates.
Earth is a rotating sphere. For a perfect calculation, we would need to use a spherical coordinate system.  In a feasibility study, we use a linear coordinate system, with the launch pad being at the origin at time $t=0$.  The Earth's center has coordinates $x=0$ and $y=-R_{_E}$, where
\be
\label{3.01}
R_{_E}=6.378 \cdot 10^6\ m
\ee
is the Earth's radius.  The equation describing the Earth's surface is
\be
\label{3.02}
x^2+\big(y+R_{_E}\big)^2=R_{_E}^2.
\ee
Any point (x,y) has an altitude given by
\be
\label{3.03}
h(x,y)=\sqrt{x^2+\big(y+R_{_E}\big)^2}-R_{_E}^2.
\ee
Notice, that the rocket may only fly within the space with positive altitude.  Typical trajectories of the rocket stages are below:
\begin{center}
\includegraphics[width=16cm]{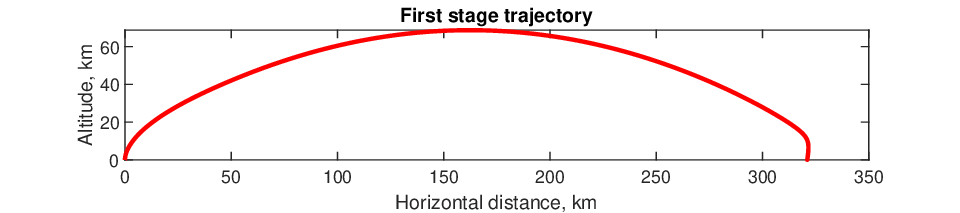}
\includegraphics[width=16cm]{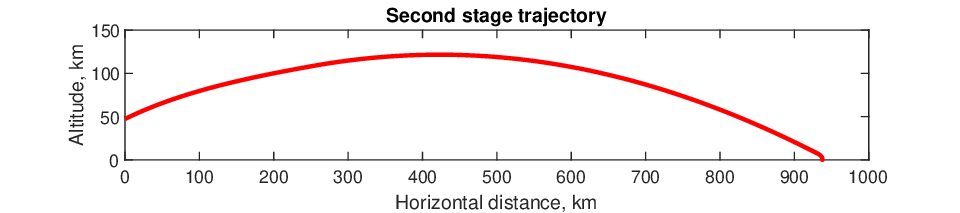}
\captionof{figure}{Typical Rocket Stage Trajectories \label{F03}}
\end{center}

In order to approximate rocket motion, we must consider Earth's rotation.  In an engineering work, we would have to introduce a $z$ -- coordinate in order to obtain a perfect result.  In a feasibility study, we use a simplification.  At a latitude of $\theta$ degrees, the speed of Earth's rotation is
\be
\label{3.04}
v_{_E}(\theta)=463.8\ m/s\ \ \cos \theta.
\ee
For the purposes of rocket flight trajectory approximation, we can approximate the Earth's rotation by rotation within xy plane with surface velocity $v_{_E}(\theta)$
\cite[p.130]{RPE}.  Under this approximation, the space or ``medium" at point (x,y) is moving at the velocity given by
\be
\label{3.05}
\mathbf{v}_m(x,y)=v_{_E}(\theta) \cdot \frac{\big(y+R_{_E}\big)\hat{x}-x \hat{y}}{R_{_E}}.
\ee
Notice, that the launching pad is moving in the direction $\hat{x}$ at velocity $v_{_E}(\theta)$ as a result of Earth's rotation.

\begin{wrapfigure}{R}{6cm}
\includegraphics[width=6cm,height=6cm]{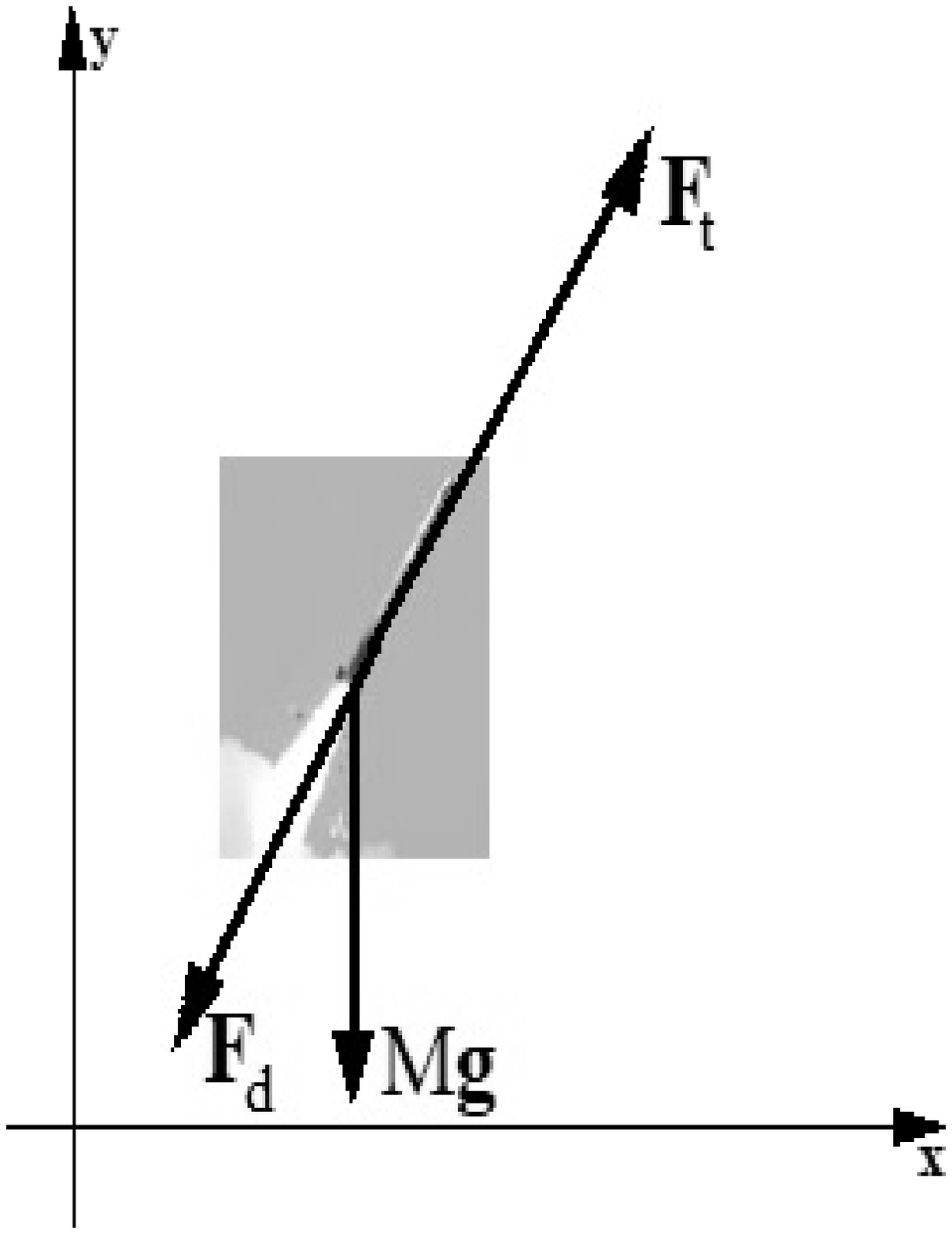}
\captionof{figure}{Forces Acting on a Firing Rocket \cite{GMLRS2} \label{F04}}
\end{wrapfigure}
Three forces acting on a rocket during ascent are presented in Figure \ref{F04}.  The gravitational force is $M(t) \mathbf{g}(x,y)$.  The mass depends on time as it is decreasing during the fuel burning.  The variable $\mathbf{g}(x,y)$ is acceleration due to gravity.  It is pointed in the direction of the Earth's center, and it is proportional to the inverse square of the distance to Earth's center.  In mathematical terms, the direction of $\mathbf{g}$ is
\be
\label{3.06}
\hat{g}(x,y)=
-\frac{x \ \hat{x}+\big(y+R_{_E}\big)\ \hat{y}}
{\sqrt{x^2+\big(y+R_{_E}\big)^2}}.
\ee
The magnitude of $\mathbf{g}$ is
\be
\label{3.07}
g(x,y)
=g_{_0}\ \frac{R_{_E}^2}{x^2+\big(y+R_{_E}\big)^2}
=g_{_0}\ \frac{R_{_E}^2}{\big(R_{_E}+h \big)^2},
\ee
where $g_{_0}=9.81\ m/s^2$, and $h$ is the altitude.  Notice, that for low altitudes and distances,
\be
\label{3.08}
g(x,y)=-\hat{y}\ g_0.
\ee
When we were working on artillery rockets and projectiles, we used (\ref{3.08}).

The air resistance or drag force $\mathbf{F}_d$ is acting in the opposite direction of velocity relative to the ``medium".  Its magnitude is
\be
\label{3.09}
F_d=C_d(\mathcal{M}) \frac{\rho v^2 A}{2}=C_d(\mathcal{M})
\ \ \mathcal{M}^2 \ \ \frac{\rho v_s^2 A}{2},
\ee
where
$C_d(\mathcal{M})$ is the Mach--number dependent drag coefficient,
$v$ is the velocity relative to the ``medium",
$v_s$ is the speed of sound,
$\mathcal{M}$ is the Mach number, i.e. $\mathcal{M}=v/v_s$,
$\rho$ is the air density,
and $A$ is the base area of the rocket.
Notice that the symbol $M$ is already taken by the rocket mass.
The thrust force is $\mathbf{F}_t$.
In dense atmosphere, it is acting in the direction of velocity.
At later stages of flight, it may act in any direction the engineers desire.
Its magnitude is
\be
\label{3.10}
F_t=\dot{M} v_e,
\ee
where $\dot{M}$ is the fuel burning rate, and $v_e$ is the exhaust velocity.  The exhaust velocity increases as the rocket rises out of dense atmospheric layers.  The fuel burning rate is
\be
\label{3.11}
\dot{M}=\frac{M\ f_p}{t_b} \qquad \text{for time} \ \ 0 \le t \le t_b,
\ee
where $t_b$ is the \textbf{burn time} and $f_p$ is the \textbf{propellant mass fraction} given by
\be
\label{3.12}
f_p=\frac{\text{Propellent mass}}{\text{Combined mass of propellant, rocket, and payload}}.
\ee

The change in velocity produced by the rocket engine is
\be
\label{3.13}
\triangle v=\int_0^{t_b} \frac{\mathbf{F}_t(t)}{M(t)}\ dt,
\ee
where $M(t)$ is the rocket mass at time $t$.
According to Tsialkovski Rocket Equation \cite{Tsialkovski},
\be
\label{3.14}
\triangle v=-\overline{v}_e\ ln \big(1-f_p \big),
\ee
where $\overline{v}_e$ is the average exhaust velocity.

If a rocket has more than one stage, then the combined $\triangle v$ of the rocket is the sum of $\triangle v$ for each stage.  The value of $\triangle v$ is very important in rocket and space flight engineering, since every maneuver has a corresponding $\triangle v$.
For Low Earth Orbit (LEO) launch, the Shuttle needs $\triangle v = 9,285\ m/s$ \cite[p. 130]{RPE}.  This is considerably greater than orbital velocity.  Such $\triangle v$ is necessary to overcome drag loss and gravitational loss described below.

In the system we are working on, $\triangle v$ is between 9,902 $m/s$ and 9,936 $m/s$, which is 617 $m/s$ to 651 $m/s$ higher than $\triangle v$ for the Shuttle.  First, the altitude of LEO given in \cite[p.130]{RPE} is 190 $km$, while our system delivers payload to a 400 $km$ orbit.  That is responsible for extra $\triangle v$ of 122 $m/s$.  Second, the drag loss of our system is 76 $m/s$ higher than that of the Shuttle.  Third, the total time during which our system performs active ascent is 12.4 $minutes$ -- which is considerably longer than ascent time for other systems.  Longer ascent times result in increased gravity loss.  In our case, the extra $\triangle v$ needed to overcome long rocket firing time is 419 $m/s$ to 453 $m/s$.  Later versions of our system with more powerful engines may have $\triangle v$ requirements lowered by about 430 $m/s$.

Producing $\triangle v = 9,285\ m/s$ and more so $\triangle v = 9,780\ m/s$ requires a very high ratio of liftoff mass to mass launched into LEO.  In Appendix A, we summarize launch systems used in 2017.  For 16 out of 29 systems, the mass placed into LEO is 1\% to 2\% of the launch mass.  For 6 systems, the mass placed into LEO is 2\% to 3\% of the launch mass.  For 5 systems, the mass placed into LEO is 3\% to 4\% of the launch mass.  For 2 systems, the mass placed into LEO is over 4\% of the launch mass. Falcon Heavy, which places 4.49\% of launch mass into LEO, is the champion.  In our four-stage rocket, the first two stages can be reused multiple times without refurbishment.  In order for them not to be over exhausted, they have combined $\triangle v = 4,434\ m/s$.  The final mass placed into LEO is 0.255\% of the launch mass.

\subsubsection{Losses of $\triangle v$}

The \textbf{drag loss} is caused by air resistance.  It is given by
\be
\label{3.15}
\triangle v_{_d}=\int_0^{t_0} \frac{F_d(t)}{M(t)}\ dt,
\ee
where $M(t)$ is the rocket mass as a function of time, $t_0$ is the time for which the drag loss is calculated, and $F_d(t)$ is the drag force at the time moment $t$.

\textbf{Gravity loss} is caused by the fact that for a significant portion of the ascent to orbit, Earth's gravitational force is acting in a direction opposite to the rocket's velocity.  It is given by
\be
\label{3.16}
\triangle v_{_g}=\int_0^{t_0} \mathbf{g}(t) \cdot \hat{v}(t) dt,
\ee
where $\mathbf{g}(t)$ is the magnitude and direction of the gravitational acceleration experienced by the rocket at time $t$, while $\hat{v}(t)$ is a unit vector in the direction of the rocket's velocity at time $t$.  For most space missions, gravity loss is the main loss of
$\triangle v$.  For Shuttle missions, gravity loss accounted for 62\% of all $\triangle v$ losses \cite[p.130]{RPE}.

\textbf{Vector loss} is caused by the fact that the rocket's vector thrust does not always point in the same direction as rocket velocity.  For a large portion of the rocket's ascent to orbit, it has to be deflected upward in order to prevent the rocket from falling into the dense atmosphere.  It is given by
\be
\label{3.17}
\triangle v_{_v}
=\int_0^{t_0} \big\| \mathbf{a}_{_r}(t)\big\|- \mathbf{a}_{_r}(t) \cdot \hat{v}(t) dt
=\int_0^{t_0} \big\| \mathbf{a}_{_r}(t)\big\| \big(1-\cos \theta \big) dt,
\ee
where $\mathbf{a}_{_r}(t)$ is the magnitude and direction of the acceleration due to rocket engines experienced by the rocket at time $t$, $\hat{v}(t)$ is a unit vector in the direction of the rocket's velocity at time $t$, and $\theta$ is the angle between the rocket velocity and rocket thrust.

\subsection{Liquid Rocket Engines}
In this section, we discuss several parameters of the engines which should be uses in the systems described in this work.  First, we present a table containing some important specifications of the proposed engines.

\subsubsection{Engine specifications}

The data for all engines used in the Orbital Delivery System is tabulated below.
In Rows 2 and 3, SL is sea level and V is vacuum.
In Row 4, NED is nozzle exit diameter.
In Row 5, EEC is the mixture of 61\% ethanolamine, 30\% ethanol, and 9\% hydrated copper nitrate.
In Row 6, LO2 is liquid oxygen, while HP95 is 95\% hydrogen peroxide.
In Row 8, OFR is the mass ratio of oxidizer to fuel.
In Row 9, CCT is the average temperature in the combustion chamber.
In Row 10, CCP is the pressure in the combustion chamber.
In Row 11, EAR is the \textbf{expansion area ratio} -- the area ratio of the nozzle exit to the nozzle throat.

\begin{center}
  \begin{tabular}{|l|l|l|l|l|l|}
     \hline
                & First           & Second        & Two stage     & One stage     \\
                & stage           & stage         & pressure fed  & pump fed      \\
                & MPDS            & MPDS          & MPTO          & MPTO          \\
     \hline
 1. Mass        & 3.0 $ton$       & 1.4 $ton$     & 600 $kg$      & 700 $kg$      \\
 2. Thrust      & 1,472 $kN$ SL   & 675 $kN$ V    & 496 $kN$      & 454 $kN$      \\
                & 1,647 $kN$ V    &               &               &               \\
 3. Exhaust $v$ & 2,530 $m/s$ SL, & 2,980 $m/s$ V & 2,813 $m/s$ V & 2,916 $m/s$ V \\
                & 2,830 $m/s$ V   &               &               &               \\
 4. NED         & 1.39 $m$        & 1.45 $m$      & 2.03 $m$      & 1.51 $m$      \\
     \hline
 5. Fuel        & Methane         & Methane       & EEC           & EEC           \\
 6. Oxidizer    & LO2             & LO2           & HP95          & HP95          \\
 7. Feed        & pump            & pump          & pressure      & pump          \\
 8. OFR         & 1.89            & 1.89          & 4.5           & 4.5           \\
 9. CCT         & 2,100 $^o$C     & 2,100 $^o$C   & 2,353 $^o$C   & 2,406 $^o$C   \\
10. CCP         & 100 $bar$       & 100 $bar$     & 25 $bar$      & 100 $bar$     \\
11. EAR         & 15              & 40            & 40            & 100           \\
     \hline
   \end{tabular}
   \captionof{table}{Engine Parameters  \label{3.T01}}
\end{center}

Many parameters such as engine mass and dimensions are still unknown.  They will have to be designed with durability and reusability being the primary considerations.

\subsubsection{Combustion chamber temperature}

For the oxidizer-fuel combinations giving high exhaust velocity, the stoichiometric combustion temperature is very high -- usually exceeding 3,300 $^o$C.  It is possible to lower the temperature within the combustion chamber in general and within any area in contact with the walls in particularly by lowering the oxidizer to fuel ratio.  This would also result in lowering of exhaust velocity, which we call \textbf{low temperature penalty}.

In the Figure \ref{F05} below, we plot the ratio of actual exhaust velocity to maximum exhaust velocity as a function of rocket nozzle throat temperature.  Both the rocket nozzle throat temperature and the exhaust velocity are functions of fuel to oxidizer ratio.  This ratio is not shown, since temperature plays the primary role.

Stage 1 engine uses methane fuel and liquid oxygen oxidizer.
The combustion chamber pressure is 100 atmospheres.
Expansion area ratio is 15.
The maximum sea level exhaust velocity is 2,930 $m/s$.
The maximum vacuum exhaust velocity is 3,210 $m/s$.

Stage 2 engine uses methane fuel and liquid oxygen oxidizer.
The combustion chamber pressure is 100 atmospheres.
Expansion area ratio is 40.
The maximum vacuum exhaust velocity is  3,390 $m/s$.

MPTO engines use 61\% ethanolamine, 30\% ethanol, and 9\% hydrated copper nitrate. fuel and 95\% hydrogen peroxide oxidizer.
Data for pressure-fed MPTO is plotted in Figure \ref{F05} below.
The combustion chamber pressure varies from 15 to 25 atmospheres.
Expansion area ratio is 40.
The maximum vacuum exhaust velocity inclusive of losses is 2,778 $m/s$.

\begin{center}
\includegraphics[width=8cm]{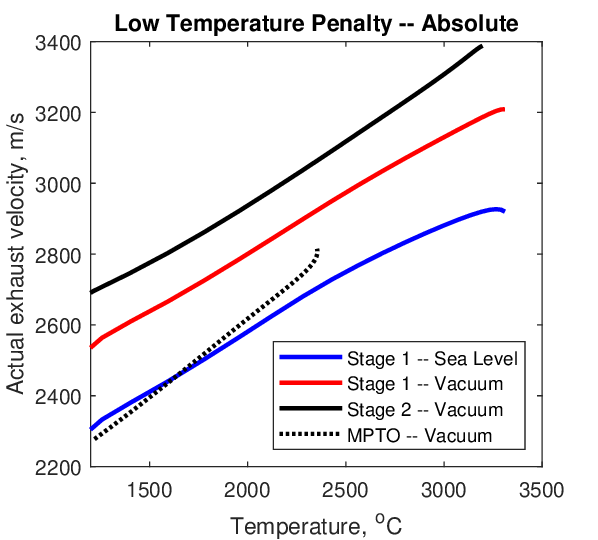}
\includegraphics[width=8cm]{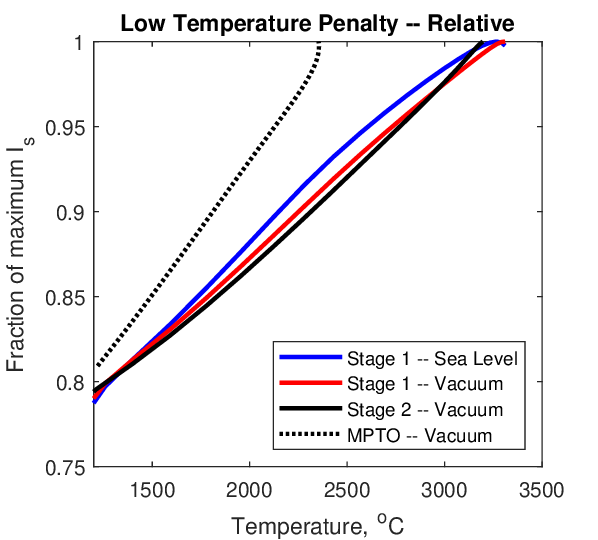}
\captionof{figure}{Low Combustion Temperature Penalty \label{F05}}
\end{center}

In MPDS engines we use oxidizer to fuel ratio producing a temperature of 2100 $^o$C.  Our exhaust velocity is 87\% to 88\% of the optimal.  Nevertheless, we must preserve the engines for hundreds or thousands of firings.

In MPTO engines we use oxidizer to fuel ratio producing the highest temperature and thrust.  First, the highest temperature produced by HP95-EEC combination is 2,406 $^o$C to 2,457 $^o$C.  Second, HP95-EEC combination has much higher low temperature penalty than liquid methane -- liquid oxygen combination.  Third, MPTO engines do not have to be reused more than about 20 times.

\subsection{Liquid Rocket Fuels and Oxidizers}

State of the art rockets use several oxidizers --
liquid oxygen,
concentrated solutions of hydrogen peroxide $\big(H_2O_2 \big)$ in water,
nitric acid $\big(H N O_3 \big)$, and
nitrogen tetroxide  $\big( N_2 O_4 \big)$  \cite[pp. 256-259]{RPE}.
State of the art rockets use several fuels --
hydrocarbon fuels,
hydrazine $\big( N_2 H_4 \big)$,
unsymmetrical dimethylhydrazine $\big( \big(C H_3 \big)_2 N N H_2 \big)$,
monomethylhydrazine $\big( \big(C H_3 \big) NH N H_2 \big)$,
\cite[pp. 259-26 ]{RPE}.
All forms of hydrazine are toxic and extremely expensive.  The aforementioned fuels and oxidizers are fit to be used by expendable rockets.  For multiply reusable rockets, some of these fuels may be useless.

\subsubsection{Cryogenic Propellant for MPDS Engines}
The engines of Stage 1 have a combined thrust of $29.4 \cdot 10^6\ N$.  These engines consume  1,370 $tons$ of propellant per flight.
The engines of Stage 2 have a combined thrust of $4.73 \cdot 10^6\ N$.  These engines consume  245 $tons$ of propellant per flight.
MPDS stages are designed to fly every day.  Stage 1 and to a lesser degree Stage 2 have voracious appetite for propellant.  The fuel-oxidizer combination should be inexpensive, and it should also deliver good exhaust velocity.

Liquid oxygen is one of the best oxidizers for the engines of large rockets.  Moreover, for a rocket which should be daily reusable, the use of any other oxidizer would be cost-prohibitive.  We have considered several fuels to be used with liquid oxygen oxidizer.

Performance of several liquid fuels is tabulated in Table \ref{4.T03} below.  In this first stage rocket engine, the combustion chamber pressure is 100 atmospheres.  The nozzle expansion area ratio is 15.
The first column lists fuels. AcetAm40 is a solution of 60\% Ammonia and 40\% Acetylene by weight.
Ethylene oxide performed so poorly, that it was not tabulated.
The second column lists the oxidizers.  LO2 is liquid oxygen.
The third column lists the oxidizer to fuel mass ratios.
The fourth column lists the average temperatures at rocket combustion chamber.  The temperatures of the flame where oxidizer and fuel contact is higher.
In the fifth and sixth columns, we consider Stage 1 rocket engine, where the combustion chamber pressure is 100 atmospheres and nozzle expansion area ratio is 15.
The fifth column lists exhaust velocities at sea level.
The sixth column lists exhaust velocities in vacuum.
In the seventh column, we consider Stage 2 rocket engine, where the combustion chamber pressure is 100 atmospheres and nozzle expansion area ratio is 40.
The seventh column lists exhaust velocities in vacuum.

The values in the last three columns are deduced from the values calculated by Rocket Propulsion Analysis (RPA) program \cite{RPA}.  These values are actual rather than ideal.  The ideal velocities are calculated by RPE.  Actual sea level exhaust velocity is the ideal sea level exhaust velocity multiplied by 0.92.   Actual vacuum exhaust velocity is the ideal vacuum exhaust velocity multiplied by 0.94.

\begin{center}
   \begin{longtable}{|l|l|l|l|l|l|l|l|l|l|l|}
     \hline
     Fuel     & Oxi-  & Oxidizer &Tempera-     & $v_e$ at 100 $atm$, & $v_e$ at 100 $atm$, & $v_e$ at 100 $atm$,\\
              & dizer & to fuel  &ture         & Expansion 15        & Expansion 15        & Expansion 40       \\
              &       & ratio    &             & sea level           & vacuum              & vacuum             \\
     \hline
     PR-1     & LO2   & 1.445    & 2,100 $^o$C & 2,370 $m/s$         & 2,648 $m/s$         & 2,793 $m/s$ \\
     PR-1     & LO2   & 1.488    & 2,200 $^o$C & 2,400 $m/s$         & 2,683 $m/s$         & 2,820 $m/s$ \\
     PR-1     & LO2   & 1.533    & 2,300 $^o$C & 2,430 $m/s$         & 2,717 $m/s$         & 2,852 $m/s$ \\
     \hline
     Ethane   & LO2   & 1.656    & 2,100 $^o$C & 2,481 $m/s$         & 2,773 $m/s$         & 2,914 $m/s$ \\
     Ethane   & LO2   & 1.709    & 2,200 $^o$C & 2,511 $m/s$         & 2,806 $m/s$         & 2,946 $m/s$ \\
     Ethane   & LO2   & 1.764    & 2,300 $^o$C & 2,542 $m/s$         & 2,842 $m/s$         & 2,980 $m/s$ \\
     \hline
     NH$_3$   & LO2   & 0.927    & 2,100 $^o$C & 2,467 $m/s$         & 2,697 $m/s$         & 2,868 $m/s$ \\
     NH$_3$   & LO2   & 0.970    & 2,200 $^o$C & 2,503 $m/s$         & 2,737 $m/s$         & 2,913 $m/s$ \\
     NH$_3$   & LO2   & 1.014    & 2,300 $^o$C & 2,537 $m/s$         & 2,774 $m/s$         & 2,955 $m/s$ \\
     \hline
     AcetAm40 & LO2   & 0.799    & 2,100 $^o$C & 2,504 $m/s$         & 2,800 $m/s$         & 2,938 $m/s$ \\
     AcetAm40 & LO2   & 0.835    & 2,200 $^o$C & 2,534 $m/s$         & 2,833 $m/s$         & 2,970 $m/s$ \\
     AcetAm40 & LO2   & 0.873    & 2,300 $^o$C & 2,565 $m/s$         & 2,868 $m/s$         & 3,003 $m/s$ \\
     \hline
     Methane  & LO2   & 1.886    & 2,100 $^o$C & 2,536 $m/s$         & 2,834 $m/s$         & 2,977 $m/s$ \\
     Methane  & LO2   & 1.948    & 2,200 $^o$C & 2,568 $m/s$         & 2,870 $m/s$         & 3,010 $m/s$ \\
     Methane  & LO2   & 2.012    & 2,300 $^o$C & 2,599 $m/s$         & 2,904 $m/s$         & 3,044 $m/s$ \\
     \hline
   \end{longtable}
   \captionof{table}{Performance of Liquid Bipropellants in First Stage Engine \label{3.T03}}
\end{center}

Liquid methane fuel the highest specific impulse.  Methane also has by far the lowest cost.  Thus, methane -- liquid oxygen is the optimal combination.

\subsubsection{Hypergolic Propellant for MPTO Engines}

\textbf{Hypergolic propellants} ignite as soon as propellant spray and fuel spray are combined.  These propellants are especially useful for MPTO, RLA, and possibly SSR engines.  Incorporating an expensive ignition system into these engines is a significant extra cost.  Most hypergolic propellants used in state of the art rockets include hydrazine or one of its derivatives, which are prohibitively expensive.

Having considered several non-cryogenic oxidizers for hypergolic combinations, we concluded that 95\% solution of hydrogen peroxide (HP95) is the best oxidizer.
An 95\% solution of hydrogen peroxide can undergo catalytic decomposition producing steam and oxygen at 870 $^o$C \cite{RPA}, which is above the autoignition temperature of most fuels.
That is an important advantage for small rockets.
In the past hydrogen peroxide had problems of instability, but with modern containers and stabilizers, concentrated hydrogen peroxide is very stable.  In drum quantities, it loses up to 0.4\% per year \cite[p. 14]{H2O2Storability}.  Dissociation rate of hydrogen peroxide is inversely proportional to water content \cite[p. 7]{H2O2Storability02} -- thus, high grade peroxide is more stable than low grade one.

Concentrated hydrogen peroxide is hypergolic with the following fuel mixtures.
First, it is hypergolic with ETA -- the mixture of ethanolamine and 10\% $CuCl_2$ \cite[p.274]{PerOx01}.
Second, it is hypergolic with ETAFA -- the mixture of 47.5\% Ethanolamine, 47.5\% Furfuryl Alcohol, and 5.0\% $CuCl_2$ \cite{PerOx02}.
Third, it is hypergolic with pyrrole \cite{PerOx03}.
Forth, it is hypergolic with EEC -- the mixture of 61\% monoethanolamine, 30\% ethanol, and 9\% hydrated copper nitrate.  On contact, EEC ignites with a delay of only
$1.6 \cdot 10^{-2}\ s$ \cite{PerOx04}.

Performance of several hypergolic combinations in MPTO engines is tabulated in Table \ref{3.T04} below.
In this third/forth stage rocket engine, the combustion chamber pressure is 10 atmospheres.  The nozzle expansion area ratio is 40.
The first column lists fuel.
EEC, ETAFA, and EEC are described in the above paragraph.
The second column lists the oxidizers.  HP95 is 95\% hydrogen peroxide.
The third column is the oxidizer to fuel mass ratio.  The fourth column is the temperature in the rocket combustion chamber.
The fifth column is the actual exhaust velocity in vacuum.  The ideal exhaust velocity is calculated by Rocket Propulsion Analysis program \cite{RPA}.  Actual vacuum exhaust velocity is the ideal exhaust velocity multiplied by 0.94.

\begin{center}
   \begin{tabular}{|l|l|l|l|l|l|l|l|l|l|l|}
     \hline
     Fuel     & Oxidizer    & Oxidizer & Combustion   & Expansion & Combustion  & Exhaust        \\
              &             & to fuel  & chamber      & ratio     & chamber     & velocity       \\
              &             & ratio    & pressure     &           & temperature &                \\
     \hline
     ETA      & HP95        &  3.8     & 25 $atm$     & 40        & 2,174 $^o$C & 2,644 $m/s$    \\
     ETAFA    & HP95        &  3.7     & 25 $atm$     & 40        & 2,395 $^o$C & 2,800 $m/s$    \\
     EEC      & HP95        &  4.5     & 25 $atm$     & 40        & 2,372 $^o$C & 2,813 $m/s$    \\
     Pyrrole  & HP95        &  5.3     & 25 $atm$     & 40        & 2,584 $^o$C & 2,940 $m/s$    \\
     \hline
     ETA      & HP95        &  3.8     & 100 $atm$    & 100       & 2,205 $^o$C & 2,735 $m/s$    \\
     ETAFA    & HP95        &  3.7     & 100 $atm$    & 100       & 2,457 $^o$C & 2,910 $m/s$    \\
     EEC      & HP95        &  4.5     & 100 $atm$    & 100       & 2,406 $^o$C & 2,917 $m/s$    \\
     Pyrrole  & HP95        &  5.3     & 100 $atm$    & 100       & 2,672 $^o$C & 3,065 $m/s$    \\
     \hline
   \end{tabular}
   \captionof{table}{Performance of Hypergolic Liquid Bipropellants  \label{3.T04}}
\end{center}

The best oxidizer-propellant combination seems to be pyrrole with 95\% hydrogen peroxide.  Nevertheless, EEC ignites within 16 milliseconds of contact with HP95  \cite{PerOx04}, while pyrrole without $Cu Cl_2$ catalyst ignites slowly.  Pyrrole with 2\% $Cu Cl_2$ catalyst ignites quickly, but it gives residue on storage \cite{PerOx03}.  Moreover, EEC-HP95 combination has much lower flame temperature.  Thus, in the MPTO engines, we use EEC-HP95 combination.

\subsection{Propellant Cost}
\subsubsection{MPDS Propellant -- \$188 Per Ton Average}
MPDS stages use methane fuel and liquid oxygen oxidizer.  One ton of propellant contains 345 $kg$ of liquid methane and 655 $kg$ of liquid oxygen.  These components are kept in separate tanks.  Their combined cost for MPDS spaceport is \$65 plus the cost of 477 $kg$ of natural gas for electric power stations.  We demonstrate this in the following paragraphs.
During the last two decades, the natural gas prices have varied between \$114 and \$608 per $ton$ with an average price of \$257 per $ton$.  Corresponding prices per ton of propellant would vary between \$119 and \$355 per $ton$ with an average price of \$184 per $ton$.  The Energy Information Administration data for wholesale natural gas prices is below \cite{NGPrices}:
\begin{center}
\includegraphics[width=16cm]{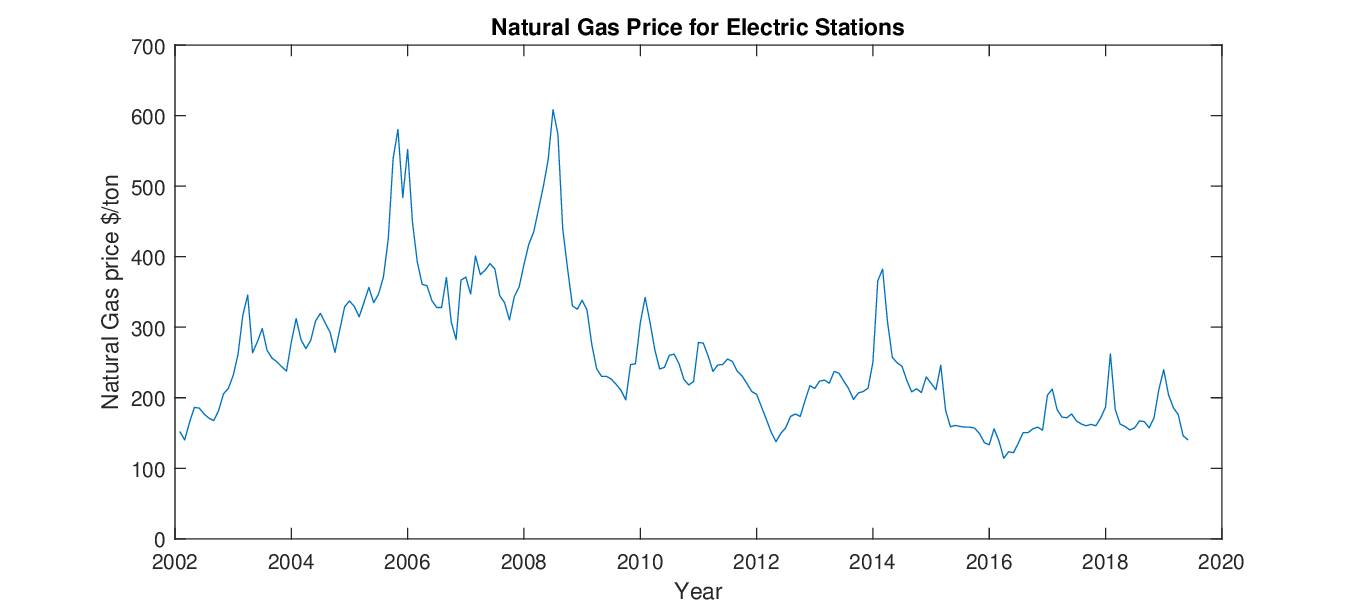}
\captionof{figure}{Natural Gas Prices \label{F06}}
\end{center}

The oxidizer used in MPDS stages is liquid oxygen.  A spaceport which regularly launches the system described in this work is not going to buy liquid oxygen.  Liquid oxygen will be generated on site by an air separation and liquefaction plant.  The same plant will liquify and purify natural gas.  The main resource needed to separate oxygen from air, liquefy oxygen, and liquefy natural gas is electricity.  Sometime in the future, the electricity will be generated by Solar Power, but for now, electricity will be generated by a natural-gas powered electric station.  Below, we calculate natural gas requirements to separate and liquefy oxygen as well as to liquefy methane.

First, we calculate electricity requirements to separate and liquefy oxygen.
It takes an average 200 $kWh/ton$ to obtain a ton of 95\% pure oxygen \cite[p.163]{AirSep1}.  This purity is insufficient for a rocket engine.  99\% pure oxygen can be obtained for 280 $kWh/ton$ \cite[p.55]{AirSep2}.
The energy cost of producing 99.5\% pure oxygen has been estimated as
320 $kWh/ton$ \cite[p.311]{AirSep3},
385 $kWh/ton$ \cite[p.22]{FishT5},
390 $kWh/ton$ \cite[p.861]{FishT3}.  We use the highest value of 390 $kWh/ton$.  Notice, that oxygen produced by separation plant is gaseous.  It takes additional energy to liquefy oxygen.  A machine working with 100\% Carnot efficiency would use 180 $kWh/ton$ \cite[p.17]{AirLiquid}, but such machines do not exist.  The best technology of 1980 liquefied oxygen for  525 $kWh/ton$ \cite[p.49]{AirLiquid}.  The state of the art now is almost the same \cite[p.155]{Cryogen}.  Using the aforementioned data, we conclude that it takes 915 $kWh$ to produce a ton of liquid oxygen out of air.  It takes 347 $kwh/ton$ to liquefy natural gas \cite[p.11]{CryoGas}.

Electric power on the spaceport is generated by natural gas burning turbine.  Some very expensive natural gas burning electric stations have efficiencies over 60\%.  A turbine of reasonable cost like General Electric (GE) LM6000 has Lower Heat Value efficiency of 41.7\% \cite[p.31]{CryoGas}.  Thus, one ton of natural gas produces 5,450 $kWh$ of mechanical or electrical energy.

As we mention in Table \ref{4.T03} above, both stages use oxidizer to fuel mass ratio of about 1.9 to 1.  Thus, each ton of propellant contains 345 $kg$ of liquid methane and 655 $kg$ of liquid oxygen.  Of course, fuel and oxidizer are stored in separate tanks.  From the data presented in the paragraphs above, generating 655 $kg$ of liquid oxygen takes
\[
655\ kg \cdot 0.915\ \frac{kWh}{kg}=600\ kWh.
\]
Liquefying 345 $kg$ of methane takes
\[
345 kg \cdot 0.347\ \frac{kWh}{kg}=120\ kWh.
\]
Overall energy requirement per ton of propellant is
\[
600\ kWh +120\ kWh=720\ kWh.
\]
We need
\[
\frac{644\ kWh}{5.45\ kWh/kg}=132\ kg
\]
of natural gas to generate that energy.  The total natural gas requirement per ton of propellant is
\[
345\ kg+ 132\ kg=477\ kg.
\]
Of the natural gas above,
\[
\frac{132\ kg}{477\ kg} \cdot 100\%=27.7\%
\]
is consumed by the power station.

In order to calculate the total cost per ton of propellant, we must also include non-fuel costs of power generation, air separation, and gas lignification.  Non-fuel costs of producing 1 $kWh$ by gas turbine power station is \$0.025 \cite[p. 26]{AirSep2}.  Thus, the non-fuel cost of producing 644 $kWh$ needed for a ton of propellant is
\[
644\ kWh \cdot \frac{\$0.025}{1\ kWh}=\$16.1.
\]
The non-capital costs of oxygen separation for a 2,300 $ton/day$ unit are \$33/ton. Air liquefaction has a non-energy cost component of \$23 per $ton$ \cite[p.16]{AirLiqCost}.  Liquid oxygen takes slightly less energy to liquefy than nitrogen.  Obtaining and liquefying 655 $kg$ oxygen has an extra cost of
\[
0.655\ ton \cdot \frac{\$33+\$23}{ton}=\$36.7.
\]
Air liquefaction has a non-energy cost component of \$36 per $ton$ \cite[p.20]{GasLiqCost}.  Liquefying 345 $kg$ methane has an extra cost of
\[
0.345\ ton \cdot \frac{\$36}{ton}=\$12.4.
\]

Adding all non-fuel costs for producing a ton of liquid propellant, we obtain
\[
\$16.1+\$36.7+\$12.4=\$65.2.
\]
The overall cost of a ton of MPDS propellant is about \$65 plus the cost of 477 $kg$ natural gas.

\subsubsection{MPTO Propellant -- \$2,000 Per Ton Average}

In the hypergolic fuel-oxidizer combinations used in MPTO rockets, the oxidizer is 95\% hydrogen peroxide.
The 50\% solution of hydrogen peroxide in water costs \$0.50 per $kg$ \cite{h2o2cost}.
According to Indian data, 50\% solution hydrogen peroxide in water costs \$0.40 per $kg$ \cite{h2o2cost2}.
Given that hydrogen peroxide purification also requires processing work, 98\% pure hydrogen peroxide should be more expensive than a similar weight of hydrogen peroxide in 50\% solution.
Hydrogen Peroxide Handbook provides the latest data from 1967 \cite{h2o2handbook}.  At that time, hydrogen peroxide cost seven times as much as now if we adjust for inflation.  An important point is that concentrating hydrogen peroxide from 70\% to 98\% increased its price on peroxide basis only by 26\%.  Thus, we can be certain that with prices for 50\% solution and modern technology, it is possible to produce 95\% pure hydrogen peroxide at \$2.00 per kg.

At this point, we calculate the cost of the fuel component of the hypergolic mixture.
Recall that
EEC   fuel consists of 61\% ethanolamine,              30\% ethanol, and 9\% hydrated copper nitrate.
ETAFA fuel consists of 47.5\% ethanolamine, 47.5\% Furfuryl Alcohol, and 5.0\% $CuCl_2$.
Furfuryl alcohol costs \$1.00 to \$2.00 per $kg$ \cite{FurfCost}.
Ethanolamine costs about \$1.80 per $kg$ \cite{EthAmCost}.
Ethanol costs \$0.50 to \$0.60 per $kg$ \cite{EthanolCost}.
Beverage alcohol is two orders of magnitude more expensive due to taxes.
The contribution of copper salts to the cost is negligible.  We conclude that both EEC and ETAFA fuels cost under \$2.00 per $kg$.

RLA may use solid propellant described in Section 1.5.  In that case, a 5,500 $kg$ grain is loaded into RLA for every flight.  The cost of propellant grain is mostly due to machining and assembly work rather than material.  Estimating the cost of that grain is beyond the scope of this work.  My educated guess is that for large production volume, the unit cost of these grains should be \$10 per $kg$.

\subsubsection{Orbital Launch Fuel Bill}

As we have shown in Section *.*, the system launching 5.1 tons of payload uses 1,615 $tons$ of MPDS propellant, which is liquid oxygen and liquid methane.  The same system uses 53 $tons$ to 56.5 $tons$ of MPTO propellant, which is EEC and HP95.  Dividing the results by 5.1 $ton$ payload, we obtain the propellant necessary to launch one $kg$ payload:\\
-- 317 $kg$ MPDS propellant;\\
-- 11 $kg$ MPTO propellant.

As we have mentioned in Subsection 2.5.2, MPTO propellant cost about \$2 per $kg$, thus MPTO fuel bill for one $kg$ payload is \$22.
In Subsection 2.5.1, we have calculated the cost of MPDS propellant based on natural gas prices over the past two decades.  MPDS propellant cost varies between \$119 and \$355 per $ton$ with an average cost of \$188 per $ton$.
Thus, MPDS fuel bill for one $kg$ payload would be \$38 to \$113 with the average of \$60.
Overall fuel bill per $kg$ payload delivered to orbit would be \$60 to \$135 with the average of \$72.

The aforementioned costs of fuel are a significant factor.  Given propellant requirements, using more expensive fuel and oxidizer would lead to a significant in crease in orbital delivery cost.  This is one more reason, why methane is the preferred fuel for MPDS.  Any MPDS oxidizer other than liquid oxygen is out of question.

For every mode of transportation which has reached maturity, fuel is a significant contributor to cost.  According to Bureau of Transportation Statistics, gasoline makes up 12\% to 33\% of average car operating cost, depending on the year \cite{Car}.  Car operating costs are plotted below:
\begin{center}
\includegraphics[width=16cm]{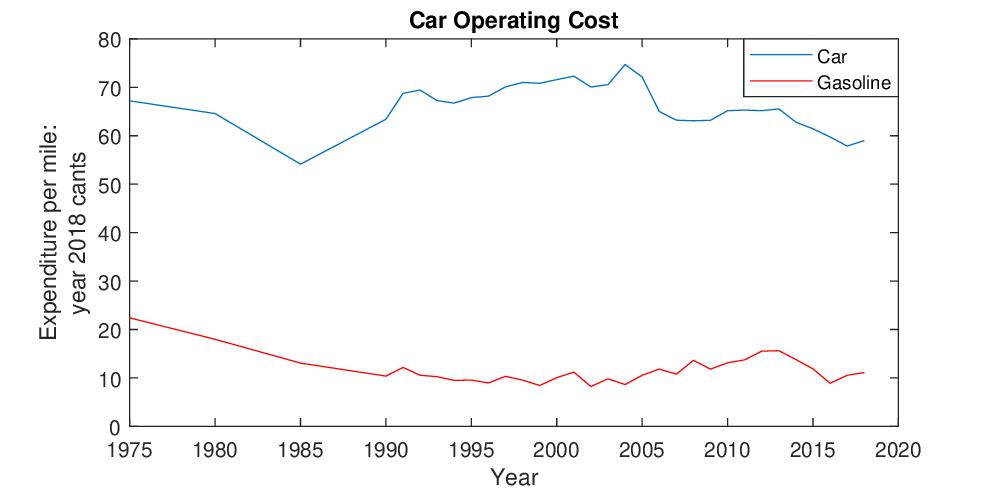}
\captionof{figure}{Car Operating Costs \label{F07}}
\end{center}
Fuel makes up 49\% of average truck operating cost \cite[p.22]{Truck}.  For cargo air carriers, fuel makes up 45\% of aircraft operating cost \cite[p.4-9]{AirLiner1}.
For an average passenger air carrier, fuel makes up 54\% of aircraft operating cost \cite[p. 4-9]{AirLiner1}.  For wide-body liners with more than 300 seats, fuel makes up 70\% of aircraft operating cost \cite[p.4-9]{AirLiner1}.

Our goal is to place payload in LEO for \$300 to \$400 per $kg$.  That would mean that 15\% to 45\% of launch system operating cost would be fuel.  The fuel cost fraction for the launch system would be greater than the one for a passenger car, but lower than the one for a truck or aircraft.

\section{Rocket Parameters and Performance}
\subsection{Midpoint Delivery System (MPDS)}
The system we are working on has three or four stages.  The first two stages should be reusable up to 400 times per year and up to 4000 times over their service life.  These stages must have wide safety margins.  These stages are called Midpoint Delivery System (MPDS).  The liftoff mass of MPDS is 2,000 $tons$.  It delivers 80 tons to a \textbf{midpoint}.

A \textbf{midpoint} is defined not only by altitude, but also by release speed and release angle.  A typical midpoint is altitude of 113 $km$, release speed of 3,000 $m/s$, and release angle of 7$^o$ to horizontal.  MPDS supplies the transport delivered to the  $\triangle v=4,400\ m/s$, about 1,400 $m/s$ of which is used to overcome gravity and air resistance.

The last two stages are simple, and their propellant tanks are disposable.  They are called Midpoint to Orbit Delivery System (MPTO).  They are described in the next chapter.

\subsubsection{Main parameters of MPDS stages}

The main parameters of the MPDS stages are listed in Table \ref{4.T01} below, and discussion follows.
In Column 5, LFM is landing fuel mass.
In Column 7, NAFMF is the mass fraction of everything except the fuel mass used in the ascent.  It is given by
\be
\label{4.01}
\text{NAFMF}=\frac{\text{Liftoff mass}-\text{Ascent fuel mass}}
{\text{Liftoff mass}}.
\ee
In column 15, SL is sea level, V is vacuum, and PA is path average.
\begin{flushleft}
   \begin{tabular}{|l|r|r|r|r|r|l|l|l|l|l|l|}
   \hline
   1.       & 4.Liftoff & 3. Stage &4. Inert & 5. LFM & 6. Ascent   & 7. NAF   & 8. Shape\\
            &   mass  & mass &    mass &              & fuel mass   &    MF    & \\
   \hline
   Stage 1  & 2,000 $ton$  & 1,600 $ton$  & 230 $ton$ & 110 $ton$ & 1,260 $ton$ & 0.37   & Cylindric fuel tank,\\
   Stage 2  &    400 $ton$ &  320 $ton$   &  75 $ton$ &  60 $ton$ &   185 $ton$ &  0.538  & engines at the end \\
   \hline
   \end{tabular}\\
   \begin{tabular}{|l|r|l|r|r|l|l|l|l|l|l|}
   \hline
            & 9. Volume     & 10. Height                 & 11. Dia-      & 12. Heat shield \\
            &               &                            & meter         &                 \\
   \hline
   Stage 1  & 2,149 $m^3$   & 17.5 $m$, 4 $m$ for engines & 12.5 $m$     & AIM162 heat sink \\
   Stage 2  &   441 $m^3$   & 10 $m$, 3 $m$ for engines   & 7.5 $m$      & AIM162 heat sink \\
   \hline
   \end{tabular}\\
   \begin{tabular}{|l|r|r|r|r|l|l|l|l|l|l|}
   \hline
            & 13. Engines            & 14. Firing & 15. Exhaust velocity, $m/s$ & 16. Stage \\
            &                        &     time   &                    & $\triangle v$ \\
   \hline
   Stage 1  & 20 $\times$ 1,472 $kN$ & 115 $s$    & 2,530 SL; 2,830 V; 2,723 PA  &
   2,708 $m/s$\\
   Stage 2  &  7 $\times$   675 $kN$ & 120 $s$    & 2,980 $m/s$ V & 1,847 $m/s$ \\
   \hline
   \end{tabular}
   \captionof{table}{The two MPDS Stages \label{4.T01}}
\end{flushleft}

Stages 1 and 2 deliver an 80 ton load to the midpoint.  Of this load, 67 $tons$ consist of Stage 3, Stage 4, and a payload, while 13 $tons$ consist of expandable dome.

\begin{wrapfigure}{R}{5cm}
\includegraphics[width=5cm,height=5cm]{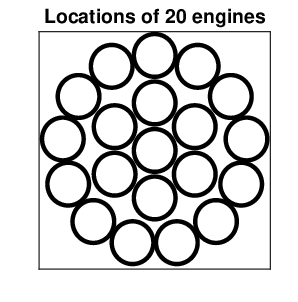}
\captionof{figure}{Locations of 20 Engines of the First Stage \label{F08}}
\end{wrapfigure}

Stage 1 is an Extra Heavy Class rocket.  The sides and especially the top of the stage are covered with a heat sink shield, which is described in Subsection 4.3 below.  Stage 1 is powered by 20 engines.  Their arrangement is presented in Figure \ref{F08}.  Each engine has a sea level thrust of 1,374 $kN$, a vacuum thrust of 1,537 $kN$, and a flight average thrust of 1,472 $kN$.  All engines use liquid methane fuel and liquid oxygen oxidizer.  Engines and fuels are discussed in Section 5.  This stage may have four movable fins on the sides.  It can be flown at least once a day.  It can fly many times between refurbishments.

Stage 2 is a Medium Class rocket.  The sides and especially the top of the stage are covered with a heat sink shield, which is much heavier than the corresponding heat sink shield on Stage 1.  Stage 2 is powered by 7 engines.  Their arrangement resembles the inner circle of Figure \ref{F08}.  Each engine has a sea level thrust of 523 $kN$, and a vacuum thrust of 675 $kN$.  They are used for vacuum ascent, vacuum deceleration and sea level landing.  All engines use liquid methane fuel and liquid oxygen oxidizer -- like Stage 1.  This stage has four movable fins on the sides, like Stage 1.  Like Stage 1, it can be flown once a day, and can work a long time without refurbishment.

\subsubsection{Performance and timetable of MPDS stages}
Performance and tentative timetables for different stages have been calculated by MatLab programs we have written.  In these programs, we have used formulas presented in Section 3.


Stage 1 works for 115 $s$.
It raises the other stages to an altitude of about 47 $km$ and releases them at about 1,542 $m/s$ at an angle of 25 $^o$ to horizontal.  After separation, Stage 1 contains 230 tons of inert mass and 110 tons of landing fuel.  The Stage 1 continues following a ballistic trajectory toward a launching pad, with its engines directed downward.  Stage 1 returns to an altitude of 40 $km$ at the speed of 1,514 $m/s$.
The greatest deceleration experienced by Stage 1 is 6.1 $g$ is.
By the time Stage 1 reaches altitude of 5 $km$, it decelerates to 524 $m/s$.  At that point it turns 180 $^o$, positioning its engines downward.  Stage 1 uses up to 110 tons of fuel for landing.  It lands 319 $km$ east of the launch pad.  Stage 1 is returned to the launch pad via a fast barge, which should take 10 hours.

Stage 2 works for 120 $s$.  It raises the other stages to an altitude of 115.2 $km$ and releases them at about 2,919 $m/s$ at an angle of 7.4 $^o$ to horizontal.
After separation, Stage 2 contains 75 tons of inert mass and 60 tons of landing fuel.
By the time Stage 2 returns to an altitude of 70 $km$, its speed is 3,062 $m/s$.
At that point, the Stage 2 fires its engines with force acting in the downward direction.
This is done in order to lower the entry angle for the stage entry and thus decrease the maximum deceleration.
During the firing, Stage 2 is slowly rotating in order to evenly distribute aerodynamic heating over the sides of the rocket.
The engines burn 27 tons of fuel, supplying $\triangle v=630\ m/s$.
Then Stage 2 turns its engines backward, exposing its most reinforced part to the oncoming wind.
Stage 2 returns to an altitude of 50 $km$ at the speed of 2,948 $m/s$.
The greatest deceleration experienced by Stage 2 is 7.8 $g$ is.
By the time Stage 2 reaches an altitude of 5 $km$, it decelerates to about 339 $m/s$.  At that point it turns 180$^o$, positioning its engines downward.  Stage 2 uses up to 16 tons of fuel for landing.  It lands 906 $km$ east of the launch pad.  Stage 2 is returned to the launch pad by a hovercraft, which should take 10 hours.  The heaviest hovercraft currently available is Zubr, which can carry up to 131 tons cargo \cite{Hovercraft}.  Developing a heavier and faster hovercraft should be much less difficult, than developing new spaceships.

\subsection{Midpoint to Orbit Delivery System (MPTO)}
In this section, we describe two MPTOs -- a pressure-fed two stage hypergolic rockets and a pump-fed hypergolic rocket.  Both a pressure-fed two stage hypergolic rockets and a pump-fed hypergolic rocket deliver a 5.1 $ton$ payload within a 600 $kg$ container to the 400 $km$ LEO.

Both types of MPTO have their advantages and disadvantages in terms of cost.
The pressure-fed two stage hypergolic rocket discards tanks and pressurization system of the first stage.  The discarded system has a mass of about 4 $tons$ and cost of about \$250,000.  Tanks and pressurization system of the second stage are worth about \$120,000, and they are not reusable either.  The non-reusable parts of pump-fed hypergolic rocket cost about \$120,000.  Overall, the pump-fed hypergolic rocket delivers the same payload to orbit at lower cost compared to the pressure-fed two stage hypergolic rockets.

The main disadvantage of the pump-fed hypergolic rocket is an expensive engine that costs \$1.2 million.  One way to overcome this disadvantage is by setting up an orbital infrastructure consisting of one or more space stations.  On these stations, the MPTO engines would be packed and loaded onto Reentry Vehicles.  These vehicles would return to Earth, delivering all MPTO engines back for multiple reuse.

\subsubsection{Pressure-fed two stage hypergolic rocket}
\begin{center}
\textbf{Physics of pressure-fed rockets}
\end{center}
In a liquid-rocket engine, both fuel and oxidizer are fed into a combustion chamber.  Both fuel and oxidizer come into the chamber through hundreds of spray nozzles in order to ensure rapid mixing.  Inside combustion chamber, fuel reacts with oxidizer producing combustion with temperature between 2,000 $^o$C and 3,500 $^o$C.  Combustion products are expanded through a nozzle producing jet stream in one direction and thrust in the opposite direction.

Both oxidizer and fuel have to be fed into the combustion chamber at a pressure of at least 15 $bar$.  There are two types of systems which pressurize fuel and oxidizer for the combustion chamber.  In a pump-feed rocket, the fuel and oxidizer come into the pump at a relatively low pressure.  The pump pressurizes fuel and oxidizer and feeds them into the combustion chamber.  The main problem for pump-fed rocket engines is their relatively high cost \cite{Pressure1}.

In the pressure-fed rocket, the fuel and oxidizer are  contained in corresponding tanks at a pressure of at least 20 $bar$.  As fuel and oxidizer are expanded, the pressure inside the tanks is sustained by an influx of helium from the pressurization system.  The pressurization system consists of a compressed helium tank and gate valves.  The helium is kept inside the tank at a pressure of at least 300 $bar$.  The gate valves allow helium to enter fuel/oxidizer tanks if and only if the pressure in these tanks falls below the designed tank pressure.

Below, we calculate the mass of the pressure-fed system.  The total mass of the fuel and oxidizer tanks is given by \cite[p.16]{tanki}
\be
\label{4.02}
M_{_{\text{tanks}}} =2 S_{_f} \ \frac{P V}{\mathcal{S}},
\ee
where $P$ is the pressure inside fuel and propellant tanks, $V$ is the combined volume of fuel and propellant tanks, $S_{_f}$ is the safety factor or the ratio of minimal burst pressure to designed tank pressure, and $\mathcal{S}$ is the
\textbf{specific yield strength} of material composing fuel and propellant tanks.  Specific strength is the quotient of tensile yield strength to density.  It is measured in the units of
\be
\label{4.03}
\frac{\text{tensile strength}}{\text{density}}=\frac{N/m^2}{kg/m^3}=\frac{J}{kg}.
\ee

The product $PV$ of the gas held in the pressurization system has to be the same as total $PV$ of liquid held inside the tanks.  Thus, the mass of the compressed helium tank is
\be
\label{4.04}
M_{_{\text{He}}} =2 S_{_f} \ \frac{P V}{\mathcal{S}_{_{\text{He}}}},
\ee
where $M_{_{\text{He}}}$ is the mass of the helium tank, and $\mathcal{S}_{_{\text{He}}}$ is the specific yield strength of the material composing the helium tank.
Assuming that fuel, propellant, and helium tanks are made of the same material and allowing for extra mass of the tubing and the rest of the system, the total mass of the tanks and pressurization system is
\be
\label{4.05}
M \approx 5 S_{_f} \ \frac{P V}{\mathcal{S}}.
\ee

In the rest of the work, we use \textbf{pressurizer mass ratio} to denote the ratio of mass of tanks and pressurization system to the mass of fuel and oxidizer.  We calculate the pressurizer mass ratio using (\ref{4.05}):
\be
\label{4.06}
\begin{split}
R_{_t}&=
\frac{\text{mass of tanks and pressurization system}}{\text{mass of fuel and oxidizer}}
=\frac{5 S_{_f} \ \frac{P V}{\mathcal{S}}}{\rho V}
=\frac{5 S_{_f} P}{\rho \mathcal{S}},
\end{split}
\ee
where $\rho$ is the average density of fuel and oxidizer inside the tanks.

Below, we calculate pressurizer mass ratios for Stages 3 and 4.
Stage 3 has the safety factor\\ $S_{_f}=1.66$,
pressure $P=30\ bar=3 \cdot 10^6 N/m^2$,
average fuel and oxidizer density $\rho=1.25 \cdot 10^3 kg/m^2$.
The pressurizer mass ratio is
\be
\label{4.07}
R_{_t}
=\frac{5 S_{_f} P}{\rho \mathcal{S}}
=\frac{2.0 \cdot 10^4\ J/kg}{\mathcal{S}}.
\ee
Stage 4 has
the safety factor $S_{_f}=1.66$,
pressure $P=20\ bar=3 \cdot 10^6 N/m^2$,
average fuel and oxidizer density $\rho=1.25 \cdot 10^3 kg/m^2$.
The pressurizer mass ratio is
\be
\label{4.08}
R_{_t}
=\frac{5 S_{_f} P}{\rho \mathcal{S}}
=\frac{1.33 \cdot 10^4\ J/kg}{\mathcal{S}}.
\ee

In order to have low pressurizer mass ratio, we need to construct propellant and helium tanks out of material with high specific yield strength.  Aluminum 6061 T6 has specific yield strength of $1.05 \cdot 10^5\ J/kg$ \cite{MatWeb} -- thus if we construct the tanks of this metal, they will be too heavy.
Duralumin  has specific yield strength of $1.87 \cdot 10^5\ J/kg$ \cite[p.A12]{Materials}.
Maraging steel  has specific yield strength of $2.07 \cdot 10^5\ J/kg$
\cite[p.A11]{Materials}.
Titanium alloy has specific yield strength of $2.40 \cdot 10^5\ J/kg$
\cite[p.A13]{Materials}.
Unfortunately, these materials are unreasonably expensive.

Tensile properties of composite materials can not be easily defined by a single specific yield strength.  Nevertheless, approximating composite material's properties by effective yield strength gives good results. Composite materials have good effective specific yield strengths, and their prices have gone down considerably \cite{Pressure1}. Inexpensive composite pressure tank technology enables significant cost savings by the use of pressure-fed rockets \cite{Pressure1}.  Composite used in Type 3 Compressed Natural Gas (CNG) tanks has effective specific yield strength of $1.8 \cdot 10^5\ J/kg$ and cost of \$35 per $kg$.  Composite used in Type 4 CNG tanks has effective specific yield strength of
$2.4 \cdot 10^5\ J/kg$ and cost of \$60 per $kg$ \cite[p.6]{CNG}.  More advanced composites with effective specific yield strength of $4.0 \cdot 10^5\ J/kg$ may soon become affordable \cite{Pressure2}.

\begin{center}
  \textbf{Parameters of MPTO Stages 3 and 4}
\end{center}

Both stages use pressure-fed rocket engines.  Both stages use \textbf{EEC fuel} and 95\% hydrogen peroxide oxidizer, which are \textbf{hypergolic}.  \textbf{EEC fuel} contains 61\% ethanolamine, 30\% ethanol, and 9\% hydrated copper nitrate. \cite{PerOx02}.  \textbf{Hypergolic} propellant-fuel mixtures ignite upon contact.

The main parameters of the MPTO stages are listed in the Table \ref{4.T02} below, and discussion follows.
In Column 6, NAFMF is the mass fraction of everything except the fuel mass used in the ascent.

\begin{flushleft}
   \begin{tabular}{|l|r|r|r|r|r|l|l|l|l|l|l|}
   \hline
   1.       &2. Liftoff &3. Stage    &4. Inert & 5. Fuel  & 6. NAF  & 7. Shape & 8. Vol.\\
            &      mass &   mass     &    mass &    mass  &    MF   &          &        \\
   \hline
   Stage 3  & 67 $ton$  &40 $ton$    & 4 $ton$ &34 $ton$  & 0.492   & Several cylindric tanks,
   & 50 $m^3$ \\
   Stage 4  & 27 $ton$  &21.3 $ton$  &3.2 $ton$&18 $ton$& 0.300   & one engine
   & 10 $m^3$ \\
   \hline
   \end{tabular}\\
   \begin{tabular}{|l|r|r|r|r|r|l|l|l|l|l|}
   \hline
            & 9. Height  & 10. Engines            & 11. Firing & 12. Exhaust & 13. Stage \\
            &            &           &     time   &    velocity & $\triangle v$ \\
   \hline
   Stage 3  & Stages contained & Main -- 496 $kN$        & 180 $s$    & 2,813 $m/s$ & 1,995 $m/s$ \\
   Stage 4  & within the dome  & Auxiliary -- 3.5 $kN$   & 180 $s$    & 2,813 $m/s$ & 3,386 $m/s$ \\
   \hline
   \end{tabular}
   \captionof{table}{Two MPTO Stages \label{4.T02}}
\end{flushleft}
Stages 3 and 4 deliver a 5.1 $ton$ payload within a 700 $kg$ container to the 400 $km$ LEO.

Stage 3 shares a 496 $kN$ main engine and 3.5 $kN$ secondary engine with Stage 4.  These engines are not discarded.  The discardable part of Stage 3 consists of two oxidizer tanks, a fuel tank, and a pressurization system.

Below, we describe Stage 3 tanks in some detail.  Stage 3 contains two oxidizer tanks, one fuel tank, and a pressurization system.  It carries 34 tons of fuel and oxidizer.  The tanks are pressurized to 30 $bar$.  The tanks are composed of "plastic liner reinforced by composite wrap around entire tank" \cite[p.8]{CNG1} -- similar to Type 4 CNG tanks, which have specific yield strength of
$2.4 \cdot 10^5\ J/kg$.  Using (\ref{4.07}), we calculate the pressurizer mass ratio:
\be
\label{4.09}
R_{_t}
=\frac{2 \cdot 10^4\ J/kg}{\mathcal{S}}
=\frac{2 \cdot 10^4\ J/kg}{2.4 \cdot 10^5\ J/kg}
=.083.
\ee
Thus, the mass of tanks and pressurization system is 2.84 $tons$ and the cost of tanks themselves should be about \$171,000.  The rest of discardable third stage consists of tubing, etc. which weighs slightly over 600 $kg$.  Another 500 $kg$ also counted in discarded mass is the fuel which remains in crevices of tanks and tubing after the stage burnout.  The total inert mass of Stage 3 is 4.0 $tons$.  The total cost of discarded system should be about \$250,000.

Stage 4 contains the aforementioned 496 $kN$ main engine and 3.5 $kN$ secondary engine.  On this stage, the main engine thrust is 248 $kN$ due to the lower feed pressure.  Like Stage 3, Stage 4 contains two oxidizer tanks, one fuel tank, and a pressurization system.
The stage carries 18 $tons$ of fuel and oxidizer.  The tanks are pressurized to 20 $bar$.  Similar to Stage 3, the tanks have specific yield strength of
$2.4 \cdot 10^5\ J/kg$.  Using (\ref{4.08}), we calculate the pressurizer mass ratio:
\be
\label{4.10}
R_{_t}
=\frac{1.33 \cdot 10^4\ J/kg}{\mathcal{S}}
=\frac{1.33 \cdot 10^4\ J/kg}{2.4 \cdot 10^5\ J/kg}
=0.056
\ee
Thus, the mass of tanks and pressurization system is 1.1 $tons$ and the cost of tanks themselves should be about \$60,000.  The total inert mass of Stage 4 includes tanks, pressurization system, tubing, and engines.  It is 3.2 $tons$.
As we explain below, this stage may be recovered.

\begin{center}
\textbf{Performance and timetable of MPTO stages}
\end{center}

Stage 3 works for 180 $s$.  As the stage burns out, the payload has the altitude of 137 $km$ and velocity of 5,098 $m/s$.  It is moving horizontally.   As mentioned before, the discardable part of Stage 3 consists of a 6 ton aggregate containing an empty propellant tank, empty fuel tank, and a pressurization system.  This part burns in the atmosphere.

Stage 4 burns 17 tons of fuel in 180 $s$.  As the stage burns out, the payload has the altitude of 120 $km$, velocity of 7,831 $m/s$, and horizontal flight direction.  Now the payload is in 120 $km$ LEO.

Recall, that MPDS released Stages 3 and 4 at 113 $km$ altitude, 2,919 $m/s$ speed, and 7.4$^o$ to horizontal angle of motion.  Stage 3 had $\triangle v=1,995\ m/s$ and Stage 4 has $\triangle v=3,386\ m/s$.  Calculating the vector loss experienced by Stage 3 and Stage 4, we obtain:
\be
\label{4.11}
v_{_{L}}=2,919\ m/s+1,995\ m/s+3,386\ m/s-7,831\ m/s=469\ m/s.
\ee
This velocity loss is caused by the fact that the engine thrust is not applied in the same direction as rocket motion.  It has to be applied slightly upward in order to prevent the rocket from falling into the dense atmosphere.  The average upward angle is 24.0$^o$.

An orbit at such a low altitude is very unstable due to air friction.  Over the next hour, Stage 4 burns 0.7 $tons$ of fuel and transfers itself to a 400 $km$ LEO.  The stage can still use the Auxiliary engine for slight maneuvers.

\begin{center}
\textbf{Orbital delivery}
\end{center}
The following items arrive at the 400 $km$ LEO.  First, we have 5.1 $tons$ payload in a 700 $kg$ container.  Second, we have the 496 $kN$ main third/forth stage engine and 3.5 $kN$ auxiliary engine.  We also have tanks capable of holding 3.7 $tons$ of EEC fuel and 15.2 $tons$ of 95\% hydrogen peroxide oxidizer.  About 200 $kg$ to 300 $kg$ fuel and oxidizer may remain on the tank walls and within tubing.

In order to understand the fate of the engines and other expensive items belonging to the fourth stage, we must look at the context of the proposed orbital delivery system.  A reusable rocket capable of delivering 5.1 $tons$ of payload of LEO every day can function only as a part of an extensive space program.  Thus, useful parts of Stage 4 from many launches can be accumulated on one or more space stations and delivered back to Earth in large packages via Reentry Vehicles.  Then these engines and some other items can be refurbished and reused.  Massive space exploration would make these items multiply reusable and would further decrease the cost of space launch.

\subsubsection{Pump-fed hypergolic rocket}

This MPTO system brings the payload from midpoint to orbit via a single-stage pump-fed rocket.  This rocket also uses EEC fuel and 95\% hydrogen peroxide oxidizer.

A pump-fed rocket has many advantages over a pressure-fed rocket.  First, the tanks do not have to be kept at high pressure.  Fuel and oxidizer tank masses can be considerably lower.  Second, a pump-fed rocket has higher combustion chamber pressure, expansion ratio, and exhaust velocity than a pressure-fed rocket.

The main disadvantage of the pump-fed rocket is a high engine cost.  The least expensive rocket engine so far is Merlin 1D.  This engine has a vacuum thrust of 981 $kN$ and a cost of \$1.6 Million \cite{Falcon3,RLSS}.  The MPTO pump-fed rocket engine has a thrust of 454 $kN$.  Its cost should be about a \$1.2 million.  The use of such an engine can be economical only if it is multiply reusable.

The main parameters of the pump-fed hypergolic MPTO are listed in the Table \ref{4.T03} below, and discussion follows.
In Column 6, NAFMF is the mass fraction of everything except the fuel mass used in the ascent.  The stage consists of several cylindrical tanks with the engine at the end.

\begin{flushleft}
   \begin{tabular}{|l|r|r|r|r|r|l|l|l|l|l|l|}
   \hline
   1.       &2. Liftoff &3. Inert & 4. Fuel     & 5. NAFMF   & 6. Vol.    \\
            &      mass &   mass  &    mass     &            &            \\
   \hline
   MPTO     & 67 $ton$  & 5.2 $ton$ & 56.3 $ton$  & 0.156      & 50 $m^3$ \\
   \hline
   \end{tabular}\\
   \begin{tabular}{|l|r|r|r|r|r|l|l|l|l|l|}
   \hline
            &  7. Engines &  8. Firing &  9. Exhaust & 10. Stage     \\
            &             &     time   &    velocity & $\triangle v$ \\
   \hline
   MPTO     &  454 $kN$   & 360 $s$    & 2,878 $m/s$ & 5,347 $m/s$   \\
   \hline
   \end{tabular}
   \captionof{table}{Pump-fed Hypergolic MPTO  \label{4.T03}}
\end{flushleft}

MPTO burns 55.3 tons of fuel in 360 $s$.  As the stage burns out, the payload has the altitude of 122 $km$, velocity of 7,835 $m/s$.  Now the payload is in 122 $km$ LEO.

Recall, that MPDS released Stages 3 at 113 $km$ altitude, 2,996 $m/s$ speed, and 7.4$^o$ to horizontal angle of motion.  Stage 3 had $\triangle v=5,347\ m/s$.  Calculating the vector loss experienced by Stage 3, we obtain:
\be
\label{4.12}
v_{_{L}}=2,919\ m/s+5,347\ m/s-7,835\ m/s=431\ m/s.
\ee
As in the case of two stages, this loss is caused by the fact that the engine thrust is not applied in the same direction as rocket motion.  It has to be applied slightly upward in order to prevent the rocket from falling into the dense atmosphere.  The average upward angle is 23$^o$.

An orbit at such a low altitude is very unstable due to air friction.  Over the next hour, MPTO burns 0.8 $tons$ of fuel and transfers itself to a 400 $km$ LEO.  Pump-fed hypergolic MPTO delivers 5.1 $ton$ payload within 600 $kg$ container to 400 $km$ LEO.  It also delivers 700 $kg$ engine, 3.6 $ton$ fuel and oxidizer tanks with tubing, and about 500 $kg$ propellant stuck within tanks and tubing.

At the 400 $km$ LEO, the engine along with tanks and tubing would be delivered to a space station.  At the station, the engine is partially disassembled and packed.  The tanks and tubing with propellant scrubbed from them may be cut up and transformed into packaging material for the engine.  The engines would be packed in sets of 50 to 100 and delivered to Earth via Reentry Vehicles.

\section{Aerodynamic Heating and Thermal Protection}
An object moving through atmosphere or any gas at supersonic velocity experiences two types of aerodynamic heating.  The first type is shock wave heating.  The supersonic object produces a shock wave in front of it.  Air/gas temperature behind a shock wave can be very high -- up to 20,000 $^o$C \cite[p. p.9]{Shields}.  For the rockets we are considering, the temperature of the shocked air will never exceed ``mere" 3,500 $^o$C, but this temperature is still high.  The second type is skin friction heating.  This is due to the high-speed friction of the rocket nose and sides against the supersonic or hypersonic stream.

In this work, we use simplified expressions to find an upper bound of heat flux.  Exact determination of heat flux requires extensive calculations using Aerodynamic Theory.  Designing an actual thermal protection requires not only advanced theoretical and computational work, but also many laboratory experiments.

The expressions themselves do not differentiate between heat flux caused by the shock wave and heat flux caused by friction.  The heat transfer equations fall into two different classes.  These classes represent two types of flow -- laminar and turbulent.

At the nose of the vehicle, the flow is always laminar.  At a short distance from the nose, it becomes turbulent and stays turbulent for the rest of the vehicle length.  Generally, turbulent flow transfers a higher heat flux to the wall than a laminar flow.

The expressions used to estimate heat fluxes on different types of surfaces have been obtained over decades as mathematical fits to experimental data.  These expressions may or may not have deeper physical meanings.  Some data patterns may have more than one expression providing a good fit.

One of the main advantages our system has over the Shuttle and more so over Single Stage to Orbit (SSTO) concepts is the relatively low heating of MPDS Stages during their return.  In Subsection ({\rr *.*}), we calculate the \textbf{heat quotient} -- which is the quotient of the total thermal energy absorbed by the rocket's heat shield and the mass of the rocket.  For Stage 1, the heat quotient is 5 $kJ/kg$.  For Stage 2, the heat quotient is 24 $kJ/kg$.  Low heat quotient allows MPTO stages to have a relatively light, yet sturdy and simple heat shields.  As we see in Eq. (\ref{5.59}), vehicles returning from orbit have heat quotients of 300 $kJ/kg$ to 1,500 $kJ/kg$.  These vehicles have to use ablative heat shields.  Shuttle and SSTO, which must have reusable heat shields, must use exotic materials and subject them to extreme stresses.  Such over-exploitation is a recipe for failure.

\subsection{Laminar Heating}
As a rocket flies through the air at high speed, it experiences aerodynamic heating.  The shock wave produced by the nose cone as well as skin friction heat the air in contact with the moving rocket.  In this subsection, we present general expressions for the heat flux transferred to  different points on a rocket surface by a laminar flow.

\textbf{Recovery temperature} is the temperature the surface would have if it lost zero energy by thermal radiation or inward conduction.  Recovery temperature is such that air at that temperature has enthalpy $H_{_{aw}}$.  Adiabatic wall enthalpy, $H_{_{aw}}$ is given by
  \be
  \label{5.01}
  H_{_{aw}}=H_{_a}+r_{_f}\ \frac{v^2}{2}.
  \ee
where $H_{_a}$ is the ambient air enthalpy, $v$ is the vehicle velocity, and $r_{_f}$ is the \textbf{recovery factor} \cite[p.9]{nose}.  The recovery factor is a dimensionless number, which is $1$ at a stagnation point, and $0.9$ for a turbulent boundary flow \cite[p.100]{TMD}.  A crude approximation for the recovery temperature is given by \cite[p.100]{TMD}:
  \be
  \label{5.02}
  T_r=T_a \left( 1+.2 r_{_f} \mathcal{M}^2 \right),
  \ee
where $T_r$ is the recovery temperature, $T_a$ is the ambient air temperature.

\subsubsection{Stagnation Point Heat Flux}

\emph{The stagnation point} is a point in a flow field where the local velocity of the fluid is zero \cite[p.17]{Aerodynamics}.  The nose tip is the only stagnation point in the flow around a rocket.  The laminar heat flux to the stagnation point can be approximated by \cite[p.6,  Eq. (40)]{nose}:
  \be
  \label{5.03}
  \dot{Q}_{_S}^{^{\text{laminar}}}=7,190\ \frac{W}{m^2}\  \sqrt{\frac{\rho}{r}}\ \ \mathcal{M}^3\ \ \
  \frac{H_{_{aw}}-H_{_w}}{H_{_{aw}}-H_{_a}},
  \ee
where $\rho$ is the air density measured in $kg/m^3$, $r$ is the nose radius in meters,
$H_{_w}$ is the specific enthalpy of the air at the wall,
$H_{_{aw}}$ is the specific enthalpy of the air at a hypothetical adiabatic wall,
$H_{_a}$ is the specific enthalpy of the ambient air.
$Q_{_S}$ denotes the total thermal energy absorbed by the stagnation point, and  $\dot{Q}_{_S}$ is its time derivative.
For a typical sound velocity of 340 $m/s$, (\ref{5.03}) can be written as:
  \be
  \label{5.04}
  \dot{Q}_{_S}^{^{\text{laminar}}}=\dot{Q}_{_{K}}\  \sqrt{\frac{\rho}{r}}\ \
  v^3 \ \frac{H_{_{aw}}-H_{_w}}{H_{_{aw}}-H_{_a}},
  \ee
where $v$ is the rocket velocity in $m/s$ with respect to air and
  \be
  \label{5.05}
  \dot{Q}_{_{K}}=1.83 \cdot 10^{-4}\ \frac{W}{m^2}
  \ee
is a constant.  Sutton and Graves give a similar expression \cite{nose1} with
  \be
  \label{5.06}
  \dot{Q}_{_{K}}=1.74 \cdot 10^{-4}\ \frac{W}{m^2}
  \ee
According to Chapman \cite{nose2},
  \be
  \label{5.07}
  \dot{Q}_{_{K}}=1.63 \cdot 10^{-4}\ \frac{W}{m^2}
  \ee
Detra and Hidalgo give a velocity dependent expression \cite{nose3},
  \be
  \label{5.08}
  \dot{Q}_{_{K}}=1.45 \cdot 10^{-4}\ \frac{W}{m^2}\ V_{_{\text{kps}}}^{0.15},
  \ee
where $V_{_{\text{kps}}}$ is the rocket velocity in kilometers per second.  In this book, we use \\ $\dot{Q}_{_{K}}=1.83 \cdot 10^{-4}\ W/m^2$, which is the highest heating rate.

In a feasibility study, an estimate accurate to 20\% is more than sufficient.  In an actual design, the heat flux at every point and under a wide range of conditions will have to be calculated with a great accuracy.

According to \cite{Shubov1}, a crude approximation to (\ref{5.04}) is
  \be
  \label{5.09}
  \dot{Q}_{_S}^{^{\text{laminar}}}=\dot{Q}_{_{K}}\  \sqrt{\frac{\rho}{r}}\ \
  v^3 \ \frac{H_{_{aw}}-H_{_w}}{H_{_{aw}}-H_{_a}}
  =\dot{Q}_{_{K}}\  \sqrt{\frac{\rho}{r}}\ \
  v^3 \ \frac{T_r-T_w}{T_r-T_a},
  \ee
where $T_w$ is the rocket wall temperature, $T_a$ is the temperature of surrounding air, and $T_r$ is the recovery temperature given in (\ref{5.01}).
By including the blackbody radiation emanating from the rocket, we obtain the total heat flux passing through the stagnation point:
  \be
  \label{5.10}
  \dot{Q}_{_S}^{^{\text{laminar}}}=\dot{Q}_{_{K}}\  \sqrt{\frac{\rho}{r}}\ \
  v^3 \ \frac{H_{_{aw}}-H_{_w}}{H_{_{aw}}-H_{_a}}-e_{_w} \sigma_B T_w^4,
  \ee
where $\sigma_B=5.67 \cdot 10^{-8}\  W/m^2 K^4$ is the Stefan--Boltzmann constant, and $e_{_w}<1$ is the wall emissivity.

\subsubsection{Semi-Round Cylinder Nose Heat Flux}

No rocket built in the past had a cylindrical or semi-round cylindrical nose.  Such nose has much higher drag coefficient than a conical nose, thus it increases a rocket's drag loss and decreases its performance.  For a multiply reusable rocket, however, a semi-round cylindrical nose makes it much easier to decelerate and land.  It also experiences less heating than a classical rocket nose.

The heat flux for a semi-round cylinder moving along its axis is given in \cite[p.428]{nose4} and plotted in Figure \ref{F09} below:
\begin{center}
\includegraphics[width=8cm]{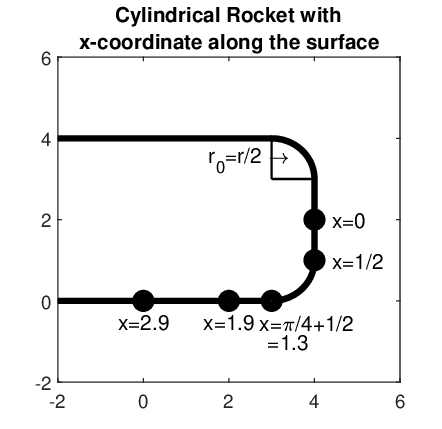}
\includegraphics[width=8cm]{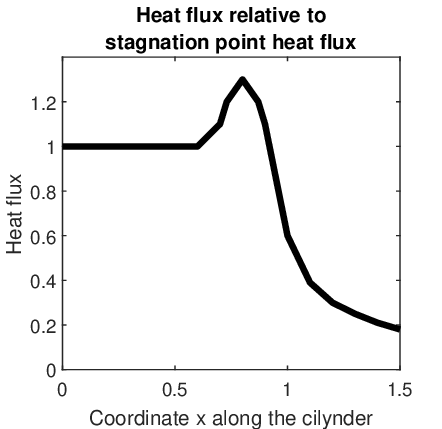}
\captionof{figure}{Heat Flux on Semi-Rounded Cylindric Rocket \label{F09}}
\end{center}
In the above plot, $x$ is a dimensionless radial coordinate along the semi-rounded cylinder surface.  For any point on the cylinder, $rx$ is the length of the shortest path to the stagnation point along the surface.  In the rest of the work, we will use the $x$-coordinate described above as the non-angular coordinate upon a semi-rounded cylindrical surface.  The angular coordinate is much less important due to cylindrical symmetry.

We calculate the total thermal energy which is absorbed by the rocket's semi-round cylindrical nose.  The energy per unit area falling upon the stagnation point during rocket flight is called \textbf{heat load}:
  \be
  \label{5.11}
  Q_{_S}^{^{\text{laminar}}}= \int \dot{Q}_{_S}(t) dt.
  \ee
Having integrated the heat flux presented in Figure \ref{F09} over the area of the cap of the rocket presented in the same figure, we obtain the total thermal energy absorbed by the rocket's semi-rounded cylindrical nose.  The nose is called "cap":
  \be
  \label{5.12}
  E_{_{\text{Hcap}}}^{^{\text{laminar}}}
  =5.0 r^2 Q_{_S}^{^{\text{laminar}}}.
  \ee

\subsubsection{End-on Cylinder Sides}

The heating of cylindrical sides of a rocket is an important factor for all rockets flying through the atmosphere at high speed.  Both stages of the Midpoint Delivery System experience significant heat flux to their cylindrical sides during return.  Second Stage Rockets experience significant heat flux to rocket sides during ascent.

For cylinder sides, the laminar heat flux is proportional to the inverse square root of the distance from the leading edge \cite[p.12]{nose5}.  The best quadratic fit based on the data presented in \cite[p.428]{nose4} gives the following expression for heat flux along the cylinder edge:
  \be
  \label{5.13}
  \frac{\dot{Q}^{^{\text{laminar}}}(r y)}{\dot{Q}_{_S}^{^{\text{laminar}}}}
  =\frac{0.12}{\sqrt{y}},
  \ee
where $ry$ as the distance from the leading edge.  The best fit on data presented in \cite[p.34]{nose7} gives a result similar to (\ref{5.13}) with slightly lower coefficient.

\subsubsection{Inclined Cylinder Sides}

Most rockets turn during their flight.  SSR generally flies head-on.  MPDS Stages pitch and yew during reentry.  Inclined cylinder experiences more laminar heat flux than given in  (\ref{5.13}).  A plane inclined by an angle $5^o \le \theta \le 25^o$ experiences an extra laminar heat flux of \cite[p.9]{nose}
  \be
  \label{5.14}
  \dot{Q}_{_{\text{plane}}}^{^{\text{Eshock}}}(ry)=
  2.42 \cdot 10^{-5}\ \frac{W}{m^2} \ \
  \sqrt{\frac{\rho}{ry}}\ \  v^{3.2}
  \big( \sin \theta \sqrt{\cos \theta} \big)
  \ \frac{H_{_{aw}}-H_{_w}}{H_{_{aw}}-H_{_a}},
  \ee
where $ry$ is the distance from the leading edge.  A cone with similar inclination experiences extra laminar heat flux of \cite[p.9]{nose}
  \be
  \label{5.15}
  \dot{Q}_{_{\text{cone}}}^{^{\text{Eshock}}}(ry)=
  4.03 \cdot 10^{-5}\ \frac{W}{m^2} \ \
  \sqrt{\frac{\rho}{ry}}\ \  v^{3.2}
  \big( \sin \theta \sqrt{\cos \theta} \big)
  \ \frac{H_{_{aw}}-H_{_w}}{H_{_{aw}}-H_{_a}},
  \ee
Using the stagnation point heating given by (\ref{5.09}), we obtain the expression for the extra heating of the inclined plane and cone by the laminar flow:
  \be
  \label{5.16}
  \begin{split}
  &\frac{\dot{Q}_{_{\text{plane}}}^{^{\text{Eshock}}}(ry)}
  {\dot{Q}_{_S}^{^{\text{laminar}}}}=
  0.63  \left(\frac{v}{2,500\ m/s} \right)^{0.2}\ \
  \frac{\sin \theta \sqrt{\cos \theta}}{\sqrt{y}},\\
  &\frac{\dot{Q}_{_{\text{cone}}}^{^{\text{Eshock}}}(ry)}
  {\dot{Q}_{_S}^{^{\text{laminar}}}}=
  1.05  \left(\frac{v}{2,500\ m/s} \right)^{0.2}\ \
  \frac{\sin \theta \sqrt{\cos \theta}}{\sqrt{y}}.
  \end{split}
  \ee
The extra heating rate for an inclined cylinder must be close to the mean of the heating rates of a plane and a cone.  Thus, we have
  \be
  \label{5.17}
  \frac{\dot{Q}_{_{\text{cylinder}}}^{^{\text{Eshock}}}(ry)}
       {\dot{Q}_{_S}^{^{\text{laminar}}}} \le
     \left(\frac{v}{2,500\ m/s} \right)^{0.2}\ \
  \frac{\sin \theta \sqrt{\cos \theta}}{\sqrt{y}}.
  \ee
For the reentry velocity under 4 $km/s$, we further simplify the above
  \be
  \label{5.18}
  \frac{\dot{Q}_{_{\text{cylinder}}}^{^{\text{Eshock}}}(ry)}
       {\dot{Q}_{_S}^{^{\text{laminar}}}} \le
  \frac{1.1\ \sin \theta}{\sqrt{y}}.
  \ee
From the above equation, we conclude that the upper bound of the heat flux upon the sides of a rocket is proportional to $1/\sqrt{y}$, where $ry$ is the distance from the leading edge.

The total laminar heat flux on pitching and yawing cylinder is obtained by adding (\ref{5.13}) and (\ref{5.18}):
  \be
  \label{5.19}
  \frac{\dot{Q}_{_{\text{cylinder}}}^{^{\text{laminar}}}(ry)}
       {\dot{Q}_{_S}^{^{\text{laminar}}}} \le
  \frac{0.12+1.1\ \sin \theta}{\sqrt{y}}.
  \ee
Transferring $\dot{Q}_{_S}^{^{\text{laminar}}}$ to the right-hand side and integrating both sides over time, we obtain the heat load:
  \be
  \label{5.20}
  Q_{_{\text{cylinder}}}^{^{\text{laminar}}}(ry) \le
  \frac{0.12+1.1\ \Big\lfloor \sin \theta \Big\rfloor }{\sqrt{y}}\   Q_{_S}^{^{\text{laminar}}},
  \ee
where $\Big\lfloor \sin \theta \Big\rfloor$ is the time-average inclination angle for a given point on the cylinder.  Given that rocket guidance will keep pitching and yawing minimized and randomized,
$\Big\lfloor \sin \theta \Big\rfloor \le 0.1$.  Thus, the total heat load on cylinder sides is
  \be
  \label{5.21}
  Q_{_{\text{cylinder}}}^{^{\text{laminar}}}(ry) \le
  \frac{0.24 }{\sqrt{y}}\   Q_{_S}^{^{\text{laminar}}}.
  \ee

\subsection{Turbulent Heating}
Turbulent flow heat flux is generally 3 to 6 times greater than laminar flow heat flux.  The hypersonic flow around any rocket almost always starts out laminar and then turns turbulent.  The transition from laminar to turbulent flow occurs at Reynolds number
$10^5 \le Re \le 10^6$ \cite[p.19]{LTT}.  The Reynolds number is defined in Equations (\ref{5.29}) and (\ref{5.31}) below.  The flow is turbulent over almost all surface area of the hypersonic vehicles we are considering in this book.

A wall parallel to the air stream and having the same temperature as the air behind the shock wave experiences the following turbulent heat flux:
  \be
  \label{5.22}
  \begin{split}
  \dot{Q}^{^{\text{turbulent}}}&=20,900\ \frac{W}{m^2}\
  \rho^{.8}
  \mathcal{M}^{2.8} (ry)^{-.2} \ \frac{H_{_{aw}}-H_{_w}}{H_{_{aw}}-H_{_a}}\\
  &=1.71 \cdot 10^{-3} \ \frac{W}{m^2}\
  \rho^{.8}
  v^{2.8} (ry)^{-.2}\ \frac{H_{_{aw}}-H_{_w}}{H_{_{aw}}-H_{_a}},
  \end{split}
  \ee
where $\mathcal{M}$ is the Mach number, $v$ is the rocket velocity in $m/s$, $\rho$ is air density in $kg/m^3$ and $ry$ is the distance from the stagnation point in $m$ \cite[p.100]{TMD}.

Below, we estimate the turbulent heat flux for general wall temperature and inclination.
According to \cite[p.iii]{nose}, the turbulent heat transfer is defined in terms of Stanton number,
  \be
  \label{5.23}
  \dot{Q}^{^{\text{turbulent}}}=
  St \cdot \rho\ v \ \Big(H_{_{aw}}-H_{_w}\Big)=
  St \cdot \rho\ v \ \Big(H_{_{aw}}-H_{_a}\Big)
  \ \frac{H_{_{aw}}-H_{_w}}{H_{_{aw}}-H_{_a}} .
  \ee
The enthalpy supplied to the air by the shock wave travelling in front of the moving rocket is equal to $v^2/2$.  Thus,
  \be
  \label{5.24}
  H_{_{aw}}-H_{_a}=\frac{v^2}{2}.
  \ee
Substituting (\ref{5.24}) into (\ref{5.23}), we obtain
  \be
  \label{5.25}
  \dot{Q}^{^{\text{turbulent}}}
  =St \cdot \rho\ v \ \big(H_{_{aw}}-H_{_a}\big)
  \ \frac{H_{_{aw}}-H_{_w}}{H_{_{aw}}-H_{_a}}
  =St \ \frac{\rho\ v^3}{2}
  \ \frac{H_{_{aw}}-H_{_w}}{H_{_{aw}}-H_{_a}}
  .
  \ee

The Stanton number at distance $x$ from the leading edge can be expressed in terms of skin friction coefficient \cite[p.305]{AHeat05}:
  \be
  \label{5.26}
  St(x)= \frac{C_f(x)/2}{1+13 \big(Pr^{2/3}-1\big) \sqrt{C_f(x)/2}},
  \ee
where $C_f(x)$ is the skin friction coefficient at distance $x$ from the leading edge and $Pr$ is the Prandtl number.  The Prandtl number for air is at least $0.71$.  The skin friction coefficient never exceeds $0.008$.  Substituting these values into (\ref{5.26}) we obtain
  \be
  \label{5.27}
  St(x) \le 0.60\ C_f(x) .
  \ee
Substituting (\ref{5.27}) into (\ref{5.25}), we obtain
  \be
  \label{5.28}
  \dot{Q}^{^{\text{turbulent}}}
  =0.30\ C_f(x) \ \rho\ v^3
  \ \frac{H_{_{aw}}-H_{_w}}{H_{_{aw}}-H_{_a}}
  .
  \ee

In order to calculate the skin friction coefficient, we introduce the \textbf{Reynolds number}.  The Reynolds number for a point on a rocket body is given by
\cite[p.6]{AHeat01}
  \be
  \label{5.29}
  Re=\frac{\rho v x}{\mu},
  \ee
where $x$ is the distance of a point from the stagnation point or the leading edge.  The dynamic viscosity of the air is \cite[p.20-21]{AirVisc}:
  \be
  \label{5.30}
  \mu \lessapprox 1.7 \cdot 10^{-5}\ \frac{kg}{m \ s}\ \left( \frac{T}{240\ ^oK} \right)^{0.7}.
  \ee
The temperature of 240 $^o$K is close to the ambient air temperature in the region where all maneuvers described in this work are performed. Combining (\ref{5.29}) and (\ref{5.30}), we obtain the Reynolds number for any point on the rocket surface:
   \be
  \label{5.31}
  Re \approx 70 \cdot 10^6
             \left(\frac{\rho}{1\ kg/m^3}\right)
             \left( \frac{v}{1\ km/s}    \right)
             \left( \frac{x}{1\ m}       \right).
  \ee
The above equation is valid for ambient air temperature for most maneuvers.  The effect of heating of air by friction and shock wave is described below.

The skin friction coefficient is very well approximated by \cite[p.4]{SkinFr02}:
   \be
   \label{5.32}
   C_f=0.295\ \frac{T_a}{T^*}\ \left[ \log \left(Re\ \frac{T_a}{T^*}\ \frac{\mu_a}{\mu^*} \right) \right]^{-2.45},
   \ee
where
$T_a$ is the ambient air temperature,
$T^*$ is the air temperature at the boundary,
$\mu_a$ is the ambient air viscosity,
and $T^*$ is the viscosity at the boundary.  Expression (\ref{5.32}) shows excellent agreement with experiment \cite[p.8]{SkinFr02}.
The temperature at the boundary is the temperature corresponding to the enthalpy $H^*$, which is \cite[p.6]{AHeat01}:
   \be
   \label{5.33}
   H^*=0.22 H_{_{aw}}+0.50 H_{_w}+ 0.28 H_{_a},
   \ee
where
$H_{_{aw}}$ is the adiabatic enthalpy,
$H_{_w}$ is the enthalpy of air at wall temperature, and
$H_{_a}$ is the enthalpy of ambient air.
The term $T_a/T^*$ in Eq. (\ref{5.32}) is due to the fact that the air density is inversely proportional to the air temperature, and the friction force is directly proportional to the air density.

From (\ref{5.30}), air viscosity grows approximately as temperature to the power 0.7.    Thus,
   \be
   \label{5.34}
   \begin{split}
   C_f&=0.295\ \frac{T_a}{T^*}\ \left[ \log \left(Re\ \frac{T_a}{T^*}\ \frac{\mu_a}{\mu^*} \right) \right]^{-2.45}
   \le 0.295\ \frac{T_a}{T^*}\ \left[ \log \left(Re\ \left(\frac{T_a}{T^*}\right)^{1.7} \right) \right]^{-2.45}\\
   &=0.295\ \frac{T_a}{T^*}\ \left[\log Re+1.7 \log \left(\frac{T_a}{T^*}\right)  \right]^{-2.45}
   =\frac{0.295\ \left[\log Re-1.7 \log \left(T^*/T_a\right)  \right]^{-2.45} }{T^*/T_a}.
   \end{split}
   \ee
In Equation (\ref{5.34}) above, $\log$ represents log base 10.  We express (\ref{5.34}) in terms of natural logarithm:
   \be
   \label{5.35}
   C_f
   =\frac{2.28\ \left[\ln Re-1.7 \ln \left(T^*/T_a\right)  \right]^{-2.45} }{T^*/T_a}
   .
   \ee
Combining (\ref{5.35}) and (\ref{5.28}), we obtain
  \be
  \label{5.36}
  \dot{Q}^{^{\text{turbulent}}}
  =\frac{0.68\ \left[\ln Re-1.7 \ln \left(T^*/T_a\right)  \right]^{-2.45} }{T^*/T_a} \ \rho\ v^3
  \ \frac{H_{_{aw}}-H_{_w}}{H_{_{aw}}-H_{_a}}.
  \ee

For the rockets and velocities we are considering, the Reynolds number varies between $1 \cdot 10^6$ and $1 \cdot 10^8$.  The quotient $T^*/T_a$ varies between 1 and 8.  For all values of $Re$ and $T^*/T_a$ considered, $C_f$ is a strongly and strictly decreasing function of $T^*/T_a$.  Hence, skin friction coefficient and turbulent heating are strongly decreasing with rising wall temperature.

\subsubsection{Rocket Cylinder}
According to figures in \cite[p.427]{nose4} and \cite[p.72]{PDistr1}, the cylinder of a rocket with semi-cylindrical cap experiences normal atmospheric pressure.
For the rocket cylinder of SSR or MPDS stages, the turbulent heat flux is calculated by (\ref{5.36}) with the Reynolds number $Re$ given by (\ref{5.31}).  In order to calculate the Reynolds number, we must define the position of the "leading edge" on the cylindrical cap.  The cap is presented in Figure \ref{F09}.

\begin{wrapfigure}{R}{8cm}
\includegraphics[width=8cm]{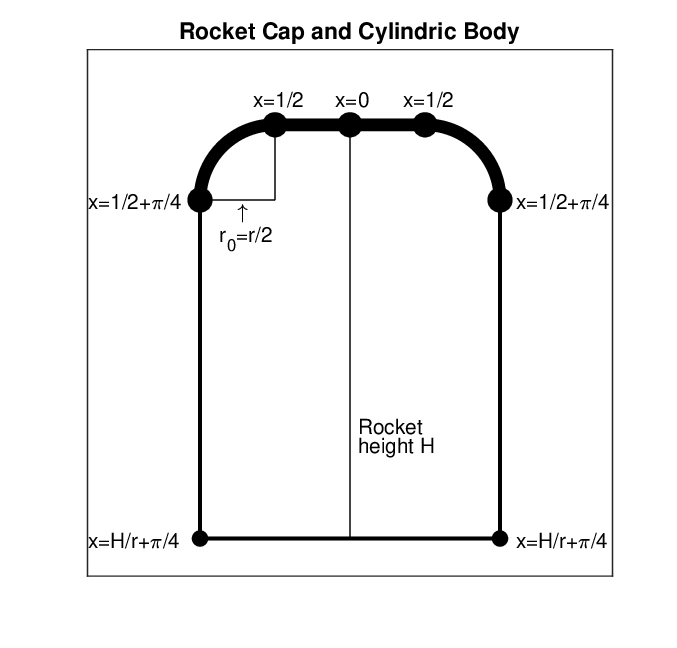}
\captionof{figure}{Rocket Cap and Body \label{F10}}
\end{wrapfigure}
In Figure \ref{F09} above, the cylinder radius is $r$.
Taking the "leading edge" $r/4$ above the body of the rocket cylinder slightly underestimates the Reynolds number and thus slightly overestimates the heating rate.  We use this value to calculate the heat flux and heat load on the cylinders of both MPDS stages.

Pressure next to an inclined cylinder side will exceed atmospheric pressure by a factor of
  \be
  \label{5.37}
  \frac{P_{_w}}{P}
  =1+\frac{\gamma}{2}\ \mathcal{M}^2\ \big(\sin \theta \big)^2
  \le 1+6\cdot 10^{-6}\ \big(\sin \theta \big)^2 v^2,
  \ee
where $\mathcal{M}$ is the Mach number and $v$ is the rocket velocity in $m/s$, and $\theta$ is the cone angle.  The heat transfer rate given in (\ref{5.36}) will be multiplied by the same amount.

SSR will not be pitching and yawing during ascent by more than 2$^o$, thus the factor given in (\ref{5.37}) will have almost no effect on SSR.  MPDS Stage 1 will reenter dense atmosphere at Mach 5 and MPDS Stage 2 will reenter dense atmosphere at Mach 10.  Rocket guidance will keep pitch and yew to minimum, nevertheless these factors may have an average effect of increasing the heat flux given in (\ref{5.36}) by factors of 1.1 for Stage 1 and 1.4 for Stage 2.

Prior to calculating the heat flux on the rocket cylinders of MPDS stages, we must include an important factor responsible for the decrease of that flux.  As a blunt object passes through the atmosphere, it drags the air along with it.  That causes the air next to the cylindrical wall to have lower velocity than the air at a distance.  The factor by which the heat flux is reduced is
  \be
  \label{5.38}
  k_{_{hf}}=\left( \frac{\mathcal{M}_w}{\mathcal{M}}\right)^3,
  \ee
where $\mathcal{M}_w$ is the Mach number of air close to the wall relative to the rocket,
and $\mathcal{M}$ is the Mach number of unshocked air relative to the rocket.

We calculate $\mathcal{M}_w$ as a function of $\mathcal{M}$.  The total energy of the air which experiences shock and drag is conserved in the rocket frame of reference.  In that frame, the air is moving toward the rocket at the velocity $\mathcal{M} v_{_s}$, where $v_{_s}$ is the ambient sound velocity.  At the stagnation point, the air slows down almost to zero, while its kinetic energy is converted to heat.  The thermal enthalpy of air at that point is
  \be
  \label{5.39}
  H_{_{\text{air}}}=\frac{\big( \mathcal{M} v_{_s} \big)^2}{2}.
  \ee
Air temperature is a function of initial temperature and extra enthalpy.  The air pressure behind the shock wave is
  \be
  \label{5.40}
  P_{_s}=P \left( 1+\frac{\gamma}{2}\ \mathcal{M}^2 \right),
  \ee
where $\mathcal{M}$ is the rocket's Mach number, and $P$ is the ambient air pressure.  As the air expands back to ambient air pressure, some of its thermal energy is transferred into kinetic energy.  We use Rocket Propulsion Analysis (RPA) software \cite{RPA} to calculate this kinetic energy.  We calculate the upper limit for the Mach number of air close to the wall relative to the rocket by
  \be
  \label{5.41}
  \mathcal{M}_w \le \frac{\sqrt{2\ KEE}}{v_{_s}},
  \ee
where $KEE$ is the kinetic energy per unit mass obtained by expansion.  Based on our RPA calculations,
  \be
  \label{5.42}
  \frac{\mathcal{M}_w}{\mathcal{M}} \le 0.82 \qquad \text{for} \qquad 3 \le \mathcal{M} \le 10.
  \ee
Given that the heat flux is proportional to the cube of velocity, the fact that air is dragged behind the rocket will multiply the heat flux on MPDS walls by a factor of
  \be
  \label{5.43}
  K_{_{\text{cylinder}}}=0.82^3=0.55,
  \ee
where $K_{_{\text{cylinder}}}$ is the cylinder heating coefficient.

In order to obtain the heat flux on MPDS Stage 1 and MPDS Stage 2 rocket cylinders, we multiply the heat flux given in (\ref{5.36}) by two factors.  The first factor due to rocket pitching and yawing during ascent is 1.1 for Stage 1 and 1.4 for Stage 2.  The second factor due to air being dragged behind the blunt rocket is 0.55 for both stages.

Thus, we use the following formulae for the maximal average heat flux on the rocket cylinders:
  \be
  \label{5.44}
  \begin{split}
  \dot{Q}^{^{\text{turbulent}}}_{_{\text{Stage 1}}}
  &=\frac{0.41\ \left[\ln Re-1.7 \ln \left(T^*/T_a\right)  \right]^{-2.45} }{T^*/T_a} \ \rho\ v^3
  \ \frac{H_{_{aw}}-H_{_w}}{H_{_{aw}}-H_{_a}},\\
  \dot{Q}^{^{\text{turbulent}}}_{_{\text{Stage 2}}}
  &=\frac{0.53\ \left[\ln Re-1.7 \ln \left(T^*/T_a\right)  \right]^{-2.45} }{T^*/T_a} \ \rho\ v^3
  \ \frac{H_{_{aw}}-H_{_w}}{H_{_{aw}}-H_{_a}}.
  \end{split}
  \ee

\subsubsection{Semi-Round Cylindrical Cap}
The semi-round caps of MPDS stages facing the air stream will experience most turbulent heating.  The heating caused by the turbulent flow is proportional to the pressure.  While the sides of MPDS stages' cylinders experience the atmospheric pressure of surrounding air, the cup experiences the pressure of the shock wave.  Recall the stagnation point pressure:
  \be
  \label{5.45}
  \frac{P_{_s}}{P}
  =1+\frac{\gamma}{2}\ \mathcal{M}^2,
  \ee
where $\mathcal{M}$ is the rocket's Mach number, and $P$ is the ambient air pressure.  Pressure over the cap is maximal at the stagnation point.  It decreases as one gets further from it.  It is shown on Figure \ref{F11} below:
\begin{center}
\includegraphics[width=16cm]{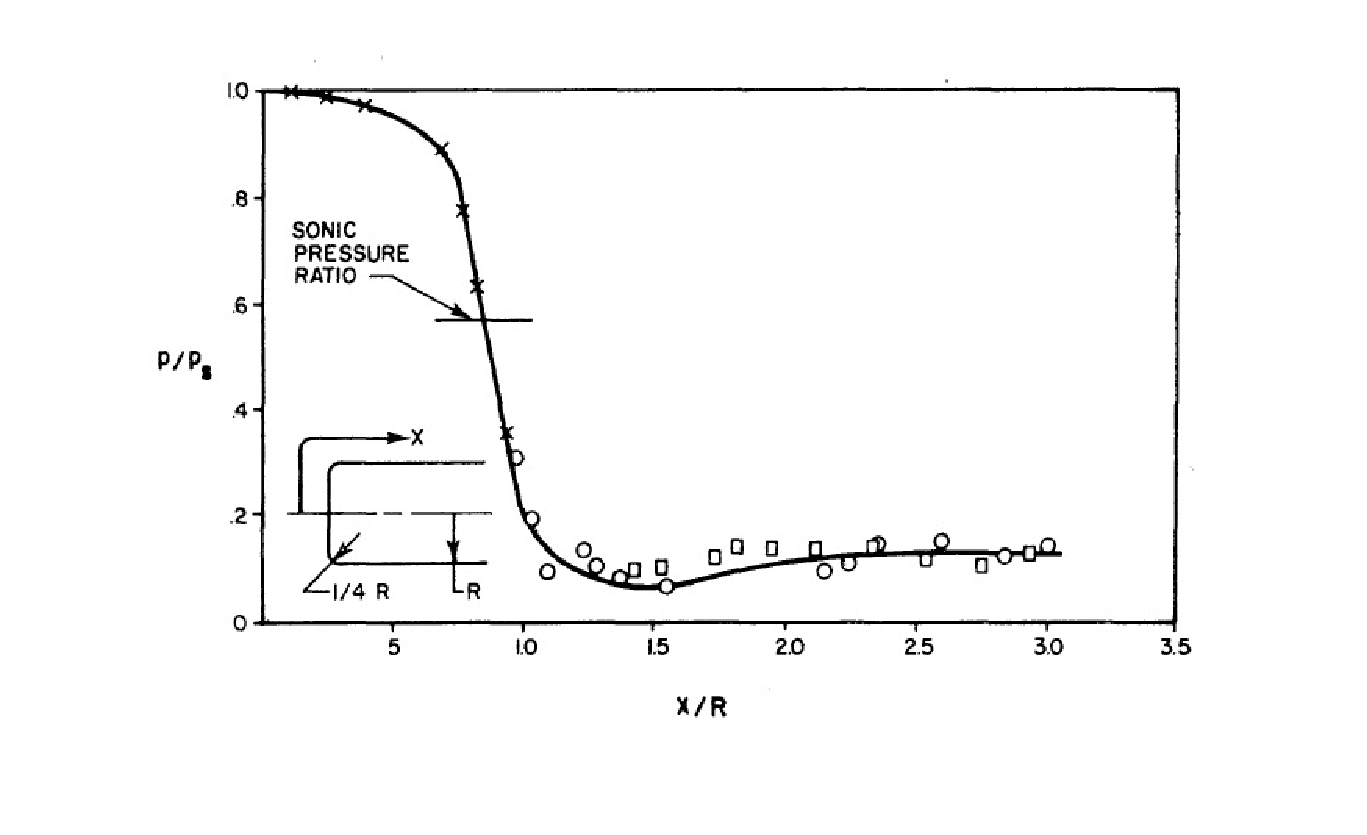}
\captionof{figure}{Aerodynamic Pressure on Semi-Cylindrical Cap \cite[p. 427]{nose4} \label{F11}}
\end{center}

From (\ref{5.36}) it follows that the friction heat flux is proportional to the cube of local air velocity.  Air velocity is zero at the stagnation point.  As the distance along the cap from the stagnation point increases, so does the local air velocity.  We describe the local air velocity increase as a function of the variable $x$ shown in Figure \ref{F09}.  As $x$ increases from 0 to $0.5$, the local Mach number increases from 0 to at most $0.3$ \cite[p.22-26]{Flow2}, \cite[p.20]{Flow3}.  At \textbf{sonic point}, the air velocity reaches the sound velocity.  In our case, as we see from Figure \ref{F10} above, the sonic point corresponds to $x=0.9$.  Notice, that the sound velocity in the air heated by the shock wave is much greater than the ambient sound velocity.  We calculate the sound velocity $v_{_{ss}}$ in air behind the shock wave below.

The temperature behind the shock wave has an upper bound of
  \be
  \label{5.46}
  T_r
  \le T \left(1+ \frac{\gamma-1}{2} \mathcal{M}^2\right)
  =T \left[1+ \frac{\gamma-1}{2} \left(\frac{v}{v_{_s}}\right)^2\right]
  ,
  \ee
where $\gamma$ is the ambient adiabatic constant of air,
$\mathcal{M}$ is the Mach number,
$v$ is the rocket velocity and
$v_{_s}$ is the ambient sound velocity.
The sound velocity behind the shock wave is
  \be
  \label{5.47}
  \begin{split}
  v_{_{ss}}&=v_{_s} \sqrt{\frac{T_r}{T}}
  =v_{_s}\sqrt{1+ \frac{\gamma-1}{2} \left(\frac{v}{v_{_s}}\right)^2}=
  \left[
  \left(\frac{\gamma-1}{2}\right)^{1/2}\sqrt{1+\frac{2}{\gamma-1}\left(\frac{v_{_s}}{v}\right)^2}
  \right] v
  \\
  &=\left[
  \left(\frac{\gamma-1}{2}\right)^{1/2}\sqrt{1+\frac{2}{\big(\gamma-1\big) \mathcal{M}^2}}
  \right] v
  \end{split}
  \ee
The pressure at the sonic point has been calculated by RPA program as
  \be
  \label{5.48}
  \frac{P_{_{sp}}}{P}
  =1+0.6 \mathcal{M}^2 \qquad \text{for} \qquad \mathcal{M} \le 10,
  \ee
where $\mathcal{M}$ is the rocket's Mach number, and $P$ is the ambient air pressure.

Having obtained the pressure and velocity at the sonic point, we can calculate the turbulent heat flux.  Using (\ref{5.36}), we obtain
  \be
  \label{5.49}
  \begin{split}
  \dot{Q}^{^{\text{turbulent}}}_{_{\text{sonic point}}}
  &=K_{_t}\ \rho_{_{\text{sonic point}}}\ v_{_{\text{sonic point}}}^3
  \ \frac{H_{_{aw}}-H_{_w}}{H_{_{aw}}-H_{_a}}
   =K_{_t}\ \left(\rho\ \frac{P_{_{sp}}}{P}\right) v_{_{ss}}^3
   \ \frac{H_{_{aw}}-H_{_w}}{H_{_{aw}}-H_{_a}}
   \\
   &=K_{_t}\ \bigg[\rho\ \left(1+0.6\ \mathcal{M}^2 \right) \bigg]
   \Bigg[ \left\{
   \left(\frac{\gamma-1}{2}\right)^{1/2}\sqrt{1+\frac{2}{\big(\gamma-1\big) \mathcal{M}^2}}
   \right\} v \Bigg]^3
   \ \frac{H_{_{aw}}-H_{_w}}{H_{_{aw}}-H_{_a}}
   \\
   &=K_{_t} \rho v^3 \cdot
   \Bigg[ \left(1+0.6\ \mathcal{M}^2 \right)
   \Bigg[  \frac{\gamma-1}{2} \cdot \left( 1+\frac{2}{\big(\gamma-1\big) \mathcal{M}^2} \right)\Bigg]^{1.5} \Bigg]
   \ \frac{H_{_{aw}}-H_{_w}}{H_{_{aw}}-H_{_a}}
   \\
   &=K_{_t}(T^*)\ K_{_{\text{sonic}}}(\mathcal{M})\ \rho v^3
   \ \frac{H_{_{aw}}-H_{_w}}{H_{_{aw}}-H_{_a}}
   .
  \end{split}
  \ee
\begin{wrapfigure}{r}{7cm}
\includegraphics[width=6cm]{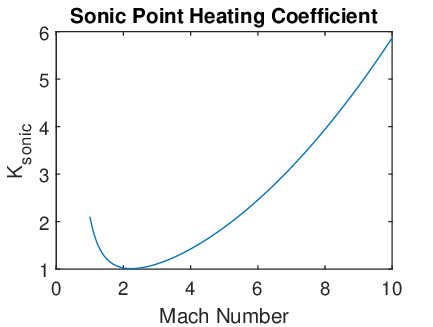}
\captionof{figure}{Sonic Point Heating \label{F12}}
\end{wrapfigure}
where $K_f$ is the coefficient given by (\ref{5.36}) as
  \be
  \label{5.50}
   K_{_t}(T^*)
  =\frac{0.68\ \left[\ln Re-1.7 \ln \left(T^*/T_a\right)  \right]^{-2.45} }{T^*/T_a},
  \ee
and the function of the sonic point heating coefficient is
  \be
  \label{5.51}
  \begin{split}
   &K_{_{\text{sonic}}}(\mathcal{M})=\\
  &\left(1+0.6\ \mathcal{M}^2 \right)
   \Bigg[  \frac{\gamma-1}{2} \cdot \left( 1+\frac{2}{\big(\gamma-1\big) \mathcal{M}^2} \right)\Bigg]^{1.5}.
   \end{split}
  \ee
The sonic point heating coefficient is plotted in Figure \ref{4.F03}.  Notice that the sonic point heating coefficient never falls below the cylinder wall heating coefficient of 0.55 given by (\ref{5.43}).

The heat load at the sonic point is obtained by integrating the heat flux:
  \be
  \label{5.52}
  Q^{^{\text{turbulent}}}_{_{\text{sonic point}}}=
  \int_t K_{_t}(T^*)\ K_{_{\text{sonic}}}(\mathcal{M})\ \rho v^3
  \ \frac{H_{_{aw}}-H_{_w}}{H_{_{aw}}-H_{_a}}
  dt.
  \ee
The friction heat flux over a blunt body is reached at the sonic point \cite[p.21]{AHeat03}.  The total area of the rocket cap shown in Figure \ref{F09} is $4.8\ r^2$.  Thus, the air friction heat which is absorbed by the nose cap during the flight is bounded by
  \be
  \label{5.53}
  E_{_{\text{Hcap}}}^{^{\text{turbulent}}}
  \le 4.8 r^2 Q^{^{\text{turbulent}}}_{_{\text{sonic point}}}.
  \ee

\subsection{Heat Sink Shields}

There are three categories of heat shields.  These are \textbf{radiative shields}, \textbf{ablative shields}, and \textbf{heat sink shields}.  \textbf{Radiative shields}, which radiate away almost all the heat flux they receive. They conduct almost no heat. \textbf{Ablative shields} absorb the heat flux.  The heat is dissipated by pyrolysis and sublimation of the shield material \cite[p.80]{Shields}.  \textbf{Heat sink shields} absorb heat flux.  Both MPDS stages use heat sink shields on their nose cap and cylindrical sides.

The heat sink shield consists of three layers.  The outermost layer is composed of metal.  The next layer is flexible thermal insulation composed of ceramic fiber.  The innermost layer is the underlying rigid structure.  The difference between heat sink shield and radiative heat shield is the function of the outer layer.  Whereas the outer layer of the radiative heat shield radiates away the heat flux, the outer layer of the heat sink shield absorbs the thermal energy.

Heat sink shields are most useful for absorbing short and powerful heat fluxes.  These shields have been used for suborbital vehicles since the 1950s \cite[p.23]{Home}.  Their main advantage is multiple reusability.  Heat sink shield is used for both MPDS stages.  The body of SSR uses a \textbf{natural heat sink shield} described in the next subsection.

The outer layer of heat sink shield is not uniform.  It consists of three sub-layers.   The outermost sub-layer is a thin sheet of refractory oxidation-resistant material such as Inconel.  The second or main layer is a metal possessing high heat capacity.  It must also have good malleability and other properties needed for any metal used in Aerospace.  The best candidate is an aluminum-beryllium alloy, AM162.  It contains 62\% beryllium and 38\% aluminum by mass \cite{AM162}.  AM162 has excellent properties.  This metal has been used in Minutman rocket \cite{AM162}.  Beryllium alloys have been used as heat sink shields for suborbital flights since the 1950s \cite[p.23]{Home}.  The third layer is also a thin Inconel sheet.

At this point, we list the properties of AM162, which make it an excellent candidate for use in a heat sink shield.  Its density is 2.1 $g/cm^3$.  AAM162's thermal conductivity is 150 $W/(m\ ^oK)$ to 200 $W/(m\ ^oK)$ in 30 $^o$C to 330 $^o$C temperature range \cite{AM162}.  The average heat capacity of the alloy between 30 $^o$C and 630 $^o$C is 1.92 $J/(g\ ^oK)$ \cite{AlHC,BeHC}.  From aforementioned data, we conclude that it takes
  \be
  \label{5.54}
  H_{_{\text{AM162}}}=1.15 \cdot 10^6 \frac{J}{kg}
  \ee
to heat AM162 alloy from 30 $^o$C to 630 $^o$C.
The heat shield will consist of 50\% AM162, 20\% Inconel and 30\% thermal insulation by mass.  Inconel has average heat capacity of 0.49 $J/(g\ ^oK)$ between 30 $^o$C and 630 $^o$C \cite[p.59]{PhysChem}.  The thermal insulation absorbs little heat.  Thus, it takes
  \be
  \label{5.55}
  H_{_{\text{HSS}}}=6.3 \cdot 10^5 \frac{J}{kg}
  \ee
to heat the heat sink shield, 30 $^o$C to 630 $^o$C.  This is the amount of heat which can be dissipated by a kilogram of heat shield material.  If the heat shield absorbs up to 1.6 times as much heat, it will overheat, but it will not break.  Thus, the heat sink shield has a wide safety margin.

The total mass of heat sink shield is the thermal energy which has to be absorbed by the heat shield divided by the energy dissipated by unit mass of heat sink shield:
  \be
  \label{5.56}
  m_{_{\text{HSS}}}=\frac{E_{_H}}{H_{_{\text{HSS}}}}.
  \ee
The areal density of the heat shield at any given point is given by dividing the heat flux at that point by the energy dissipated by unit mass of heat sink shield:
  \be
  \label{5.57}
  \sigma_{_{\text{HSS}}}=\frac{Q}{H_{_{\text{HSS}}}}.
  \ee

\subsection{MPDS Stages Reentry Heating and Deceleration}
In this subsection, calculate the total thermal energy $E_{_H}$, which must be absorbed by MPDS Stages' heat shields.  First, we calculate the thermal energy $E_{_H}^{^{\text{body}}}$ absorbed by MPDS Stages' cylindrical bodies.  Second, we calculate the thermal energy $E_{_H}^{^{\text{cap}}}$ absorbed by MPDS Stages' caps.  Third, we calculate areal densities and masses of heat shields.

The heat load on the rocket body is the sum of the heat load generated by the shock and the heat load generated by air friction.
Air friction heat load is calculated by integrating (\ref{5.44}) over time.
Shock heat load is calculated by (\ref{5.21}).
The aforementioned heat loads are calculated by the program \textbf{FlightAnalyze.m}, which is described in Appendix 4.  The heat load is calculated at several points on the cylinder of each rocket and the average heat load is obtained.  Using the average heat load, we obtain the average areal density of the heat shield by (\ref{5.57}).  We also calculate the thermal energy absorbed by the rocket's body cylinder:
  \be
  \label{5.58}
   E_{_H}^{^{\text{body}}}=2 \pi H r\ Q_{_{\text{average}}}^{^{\text{body}}},
  \ee
where $Q_{_{\text{average}}}^{^{\text{body}}}$ is the average heat load on the rocket body, $r$ is the rocket radius, and $H$ is the rocket height.

The total thermal energy absorbed by the rocket cap $E_{_H}^{^{\text{cap}}}$ is the sum of the thermal energy generated by the shock and the thermal energy generated by air friction.
Air friction thermal energy is calculated by (\ref{5.53}).
Shock thermal energy is calculated by (\ref{5.12}).
The average heat load is obtained by dividing $E_{_H}^{^{\text{cap}}}$ by the cap's area, which is $4.8\ r^2$.
Once again, all the calculation is done by \textbf{FlightAnalyze.m}.

Having obtained the total thermal energy, we calculate the heat sink shield mass, calculated by (\ref{5.56}).
Having obtained the heat flux, we calculate shield areal density by (\ref{5.57}).

The heat shield requirements for reentry of MPDS stages are presented in the left-hand side of
Table \ref{5.T01} below.
Row 1 indicates the total thermal energy absorbed by the heat shield.
Row 2 indicates required heat sink shield mass.
Row 3 indicates average heat load on the rocket cap.
Row 4 indicates required heat shield areal density over the rocket cap.
Row 5 indicates average heat load on the rocket cylindrical body.
Row 6 indicates the average heat shield areal density over the rocket cylindrical body.

Mechanical flight parameters for reentry of MPDS stages are presented in the right-hand side of
Table \ref{5.T01} below.
Row 7 indicates velocity at 40 $km$ altitude.
Row 8 indicates maximum velocity during reentry.
Row 9 indicates maximum deceleration during reentry.
Row 10 indicates rocket kinetic energy at 40 $km$ altitude.
Row 11 indicates the percentage of rocket kinetic energy turned into heat shield thermal energy during reentry.
Row 12 indicates the rocket's inert mass.
Row 13 indicates the rocket \textbf{heat quotient} -- total thermal energy absorbed by the rocket's heat shield divided by the rocket's inert mass.
\begin{center}
\begin{tabular}{|l|l|l|l|l|l|}
  \hline
                      & Stage 1        & Stage 2           \\
                      & reentry        & reentry           \\
  \hline
1. $E_{_H}$           &  0.94 $GJ$     &  1.25 $GJ$        \\
2. $M_{_{\text{HS}}}$ &  2.9 $ton$     &  3.6 $ton$        \\
3. $Q_{_{\text{average}}}^{^{\text{cap}}}$
                      &  2.5 $MJ/m^2$  &  14 $MJ/m^2$     \\
4. Cap HSAD           &  7.6 $kg/m^2$  &  39 $kg/m^2$      \\

5. $Q_{_{\text{average}}}^{^{\text{body}}}$
                      &  1.0 $MJ/m^2$  &  2.7 $MJ/m^2$     \\
6. Body HSAD          &  3.2 $kg/m^2$  &  7.6 $kg/m^2$     \\
    &            &               \\
\hline
\end{tabular}
\begin{tabular}{|l|l|l|l|l|l|}
  \hline
                   & Stage 1        & Stage 2           \\
                   & reentry        & reentry           \\
\hline
7. $v$ at 40 $km$  & 1,514 $m/s$    & 2,902 $m/s$       \\
8. Maximum $v$     & 1,541 $m/s$    & 2,948 $m/s$       \\
9. Maximum $a$     & 6.1 $g$        &  7.8 $g$          \\
10. KE40           & 404 $GJ$       & 465 $GJ$          \\
11. HS Fraction    & 0.23\%         & 0.27\%            \\
12. $M_{_{\text{inert}}}$
                   & 230 $ton$      & 75 $ton$          \\
13. $E_{_H}/M_{_{\text{inert}}}$
                   & 5 $kJ/kg$      & 24 $kJ/kg$        \\
\hline
\end{tabular}
\captionof{table}{Reentry Conditions Heat Sink Shield Requirements \label{5.T01}}
\end{center}

One noteworthy aspect of Stage 1 and Stage 2 reentries is very low heat load compared to reentries of other spacecraft.  One obvious reason is that for most atmospheric reentries, the heat load is proportional to the square of initial speed.  Suborbital reentries described in the Table \ref{5.T01} above have initial speeds of 1.6 $km/s$ and 3.0 $km/s$.  Orbital entries have initial speeds of at least 7.8 $km/s$.  Another factor is that the percentage of rocket kinetic energy turned into heat shield thermal energy during reentry is unusually low.  As we see in Table \ref{5.T01} above, in reentries of Stages 1 and 2, only 0.23\% to 0.27\% of the vehicle's initial kinetic energy is transferred to the heat shield.  In most reentries, 1\% to 5\% of the vehicle's initial kinetic energy is transferred to the heat shield \cite[p.2]{Shields}.  The heat quotient for vehicles returning from orbit is
  \be
  \label{5.59}
  \frac{E_{_H}}{M_{_{\text{inert}}}} \approx
  \frac{\left( 7,800\ \frac{m}{s}\right)^2}{2}
  \cdot \left\{ 0.01\ \text{to}\ 0.05\right\}
  = 300\ \frac{kJ}{kg}\ \ \text{to}\ \ 1,500\ \frac{kJ}{kg}.
  \ee

Both MPDS Stages have extremely blunt shapes, thus the primary fraction of the heat flux they experience is due to the shock wave rather than skin friction.  Overall heat loads upon MPDS Stages by the shock wave are relatively low.  Low shock wave heating is explained by two factors.  First, both stages are relatively large.  Second, both stages lose the bulk of their kinetic energy in relatively dense air.  As we see in Figure \ref{F13} below, both stages lose the bulk of their kinetic energy between 30 $km$ and 10 $km$.
\begin{center}
\includegraphics[width=8cm]{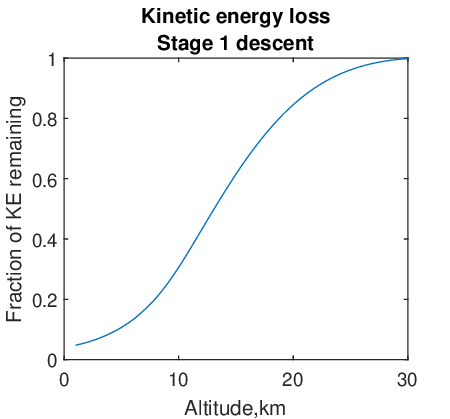}
\includegraphics[width=8cm]{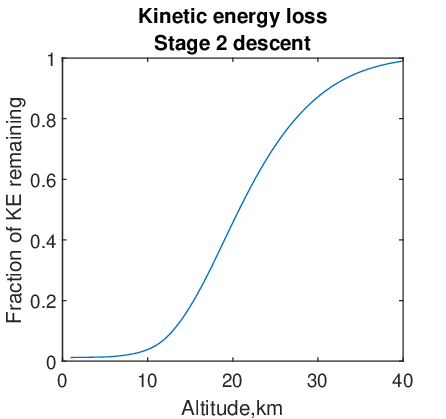}
\captionof{figure}{Kinetic Energy Loss During Descent \label{F13}}
\end{center}
Most orbital reentry vehicles lose the bulk of their kinetic energy at much higher altitudes, like 45 $km$ to 50 $km$ \cite[p.81]{Shields}.

It must be explained why large size and high air density protect the heat shield from large heat fluxes.  According to (\ref{5.09}), the stagnation point heat flux is proportional to
  \be
  \label{5.60}
  \dot{Q}_{_S} \sim \sqrt{\frac{\rho}{r}}\ \
  v^3 ,
  \ee
where $\rho$ is the air density, $r$ is the rocket radius, and $v$ is the rocket velocity.
Given that the area is proportional to $r^2$, the total heating power is proportional to
  \be
  \label{5.61}
  \dot{E}_{_H} \sim \rho^{0.5}\ r^{1.5}\ \ v^3.
  \ee
The kinetic energy loss is proportional to
  \be
  \label{5.62}
  \dot{E}_{_K} \sim \rho\ r^2\ \ v^3.
  \ee
Dividing (\ref{5.61}) by (\ref{5.62}), we obtain
  \be
  \label{5.63}
  \frac{d E_{_H}}{d E_{_K}} \sim \frac{1}{\sqrt{\rho r}}.
  \ee
The above equation shows that large objects slowing down in dense atmosphere experience much lower heat load than small objects slowing down in sparse atmosphere.

\subsection{Space Shuttle Thermal Protection}

As we have mentioned earlier, the main advantage of MPDS -- MPTO system is the fact that every stage bears relatively light load.  The main factor enabling the heat shield to function reliably is low heat quotient -- which is the quotient of the total thermal energy absorbed by the rocket's heat shield and the mass of the rocket.  The main disadvantage of the Space Shuttle is the fact that the main stage has to bear very high load in every aspect.

The heat shield of the Space Shuttle has to withstand severe heat load during orbital return.  It also has to be light.  As we have mentioned in Subsection 1.4.3, every Space Shuttle launch cost an average of \$1.54 Billion.  The Shuttle weighs 78 $tons$, and it can deliver 27.5 $tons$ to LEO \cite{Shuttle1}. From these figures, it follows, that every kilogram added to the Shuttle Heat shield would bring a recurring effective cost of \$56,000.

The Space Shuttle uses a ceramic heat shield.  It covers an area of 1,105 $m^2$ and weighs 8,574 $kg$.  The toughest part of the shield consists of reinforced carbon-carbon covering 38 $m^2$ \cite{ShuttleHP1}.  It can withstand reentry temperatures of 1,770 $^o$C for one mission and 1,630 $^o$C for 100 missions \cite{SHS01}.  The moderate part consists of LI-2200 silica tiles covering 480 $m^2$ \cite{ShuttleHP1}.  It can withstand reentry temperatures of 1,590 $^o$C for one mission and 1,260 $^o$C for 100 missions \cite{SHS01}.  Tile protection costs about \$110,000 per square meter \cite{ShuttleHP3}.  The rest of the shuttle is covered with less refractory and less expensive tiles.  Each tile requires 20 worker-hours to manufacture, test, and install.  Shuttle heat protection consists of 31,000 tiles \cite{ShuttleHP1}.

The Shuttle could have been covered by Inconel corrugated sheet heat shield instead of ceramic one. Metal heat shield is much more reliable than ceramic one.  Unfortunately, the Inconel heat shield is twice as heavy as ceramic one \cite[p.56]{NiChr01}.  Niobium heat shield capable of withstanding repeated 1,320 $^o$C entries could be produced for \$54,000 per square meter in 2018 dollars \cite[p.54]{NiobiumHS}.  It was as heavy as the Inconel shield \cite[p.52]{NiobiumHS}, thus the concept was abandoned.

On February 1, 2003, Space Shuttle Columbia suffered a disaster and seven astronauts lost their lives.  The accident was caused by failure of a single tile during reentry.  The failure triggered a chain effect \cite{ShuttleHP1}.

\section{Conclusion and Further Considerations}

\subsection{Conclusions}
The first successful space launch took place on October 4, 1957 -- a Soviet satellite named Sputnik was placed in orbit \cite{Sputnik}.  In 1961, the first astronaut named Yuri Gagarin went to space \cite{Gagarin}.  American Lunar Expedition took place in 1968.  Hopes were high for the beginning of the Space Age.

The Space Shuttle held a promise of inexpensive reusable space launch.  Then the Space Shuttle Program turned out to be 20 times more expensive than planned.  No significant improvements in Space Transportation took place between 1970 and mid 2000s.  There were many promises of reusable spacecraft which did not materialize.  The hopes for a Space Age dimmed.

By 2010, the Falcon 9 rocket by SpaceX demonstrated a significant cost reduction compared to other orbital delivery systems.  On December 21, 2015, SpaceX landed the first stage of Falcon 9 spacecraft on a launching pad.  That day, Stage 1 of Falcon 9 made a large suborbital step, which was also a great step for Humankind.

In this work, we have demonstrated the feasibility of a launch system in which the first two stages are reusable, and the third stage engine can be returned from orbit along with similar engines.  The first two reusable stages are called Midpoint Delivery System (MPDS).  The last one or two stages are called Midpoint to Orbit Delivery System (MPTO).

The two MPDS stages have relatively mild requirements in terms of $\triangle v$.  MPDS stages have wide safety margins.  Stage 1 and even Stage 2 would experience moderate but not severe aerodynamic heating.  They can be reused daily for many years -- like a commercial air liner.

The launch system is presented in Figure \ref{F02}.  The Length to Diameter (L/D) ratio for 1.4 for Stage 1 and 1.33 for Stage 2.  Most space rocket stages have much higher L/D.  This helps the non-returning rocket minimize drag loss.  High L/D of MPDS stages helps them break their velocity on return.  It also helps them during landing.  MPDS is different from conventional discardable rockets not only in function but also in shape.

MPDS uses liquid methane fuel and liquid oxygen oxidizer.  Liquid oxygen is generated from air and liquid methane is generated from natural gas.  Both of these steps are performed by electricity generated by a natural gas turbine.  MPTO uses EEC fuel and 95\% hydrogen peroxide oxidizer, which is a hypergolic mixture.  Both components of MPTO propellant cost about \$2 per $kg$.  It takes 147 tons of natural gas and 11 tons of MPTO propellant to launch a ton of payload into space.  We hope the new technology will enable space launch at \$300 per $kg$ payload.

\subsection{Further Possibilities}
The first step of vertical landing of a first stage of a space rocket has been accomplished.  At this point the challenge is to build a spaceship which is designed for reuse like an air liner.  The system described in this article is a good candidate.

Producing an accurate design of a launch system is very difficult.  Even a moderate non-reusable rocket takes 44,000 expert work-years to design \cite[p.116]{SLVD}.  Producing a launch system as reliable as a jet airliner is a monumental task, but it is feasible.

After the system is working, it can undergo some improvements.  These improvements must not come at any cost in durability and reusability.  First, we can make more powerful rocket engines for both stages.  As I mention in Subsection 3.1, engines producing higher thrust will shorten the orbital ascent time and thus reduce the total $\triangle v$ for all stages by about 430 $m/s$.  That would increase the payload about 1.3 times.  Second, we can increase engine working pressure and perhaps chamber temperature as long as it does not affect reusability.  That would increase exhaust velocity by up to 7\%.  That would further increase payload 1.45 times.  Third, every form of experience in operating the new system would lead to it's cost reduction.  Wide adoption of reusable space launch vehicles should lead to launch cost decrease to \$300 per $kg$.

\appendix

\section{Existing Launch Systems}
In the Table \ref{6.T01} below we list the launch masses and LEO payloads for the launch systems in use in 2018 \cite{Raport02}.  The first column is the launch vehicle name.  The second column is the nation launching the rocket.  The third column is the quotient of payload delivered to LEO launch mass to the launch mass.  The fourth column is the payload delivered to LEO.   The fifth column is the cost per $kg$ payload delivered to LEO.  The sixth column is the number and type of stages.  LO2K is liquid oxygen -- kerosene;  NU is N$_2$O$_4$/UDMH; LOLH -- liquid oxygen/liquid hydrogen.
\begin{center}
   \begin{tabular}{|l|l|l|r|r|l|l|l|l|}
     \hline
     Launch         & Nation    & LEO   & LEO           & SCost,           & Stages\\
     Vehicle        &           & Ratio & Payload, $kg$ & \$ per $kg$  &   \\
     \hline
     Alpha 1.0      & USA       & 1.85\% &  1,000   & 10,000   & 2 LO2K\\
     Angara A5      & Russia    & 3.10\% & 24,000   &  4,200   & 3 LO2K, 1 NU\\
     Antares        & USA       & 1.17\% &  6,200   & 12,900   & 1 LO2K, 1 Solid, 1 NU\\
     Ariane 5       & EU        & 2.56\% & 20,000   &  8,900   & 1 Solid, 2 LOLH\\
     Atlas V        & USA       & 3.31\% & 18,814   &  9,500   & 1 Solid, 1 LO2K, 1 LOLH\\
     Delta IV H     & USA       & 3.93\% & 28,790   & 13,900   & 1 Solid, 2 LOLH \\
     Dnepr          & Russia    & 1.59\% &  3,200   &  9,060   & 3 NU \\
     Electron       & USA       & 2.14\% &    225   & 21,800   & 2 LO2K \\
     Falcon 9       & USA       & 4.15\% & 22,800   &  2,720   & 2 LO2K \\
     Falcon Heavy   & USA       & 4.49\% & 63,800   &  1,410   & 2 LO2K \\
     GSLV           & India     & 1.21\% &  5,000   &  9,400   & 1 NU, 1 Solid, 1 NU, 1 LOLH\\
     Intrepid 1     & USA       & 1.55\% &    376   & 14,400   & 2 Hybrid \\
     Kuaizhou 11    & China     & 1.92\% &  1,500   & 10,000   & 3 Solid, 1 Liquid \\
     Long March 2C  & China     & 1.65\% &  3,850   &  7,790   & 2 NU, 1 Solid \\
     Long March 3A  & China     & 3.53\% &  8,500   &  8,240   & 2 NU, 1 LOLH \\
     Long March 3B/E& China     & 2.61\% & 12,000   &  5,830   & 4 NU \\
     Long March 4B  & China     & 1.69\% &  4,200   &  7,140   & 3 NU \\
     LVM3           & India     & 1.25\% &  8,000   &  7,500   & 1 Solid, 1 NU, 1 LOLH\\
     Minotaur I     & USA       & 1.60\% &    580   & 69,000   & 4 Solid \\
     Minotaur IV    & USA       & 1.85\% &  1,600   & 28,750   & 4 Solid \\
     Minotaur VI    & USA       & 2.91\% &  2,600   & 23,100   & 5 Solid \\
     Proton M       & Russia    & 3.26\% & 23,000   &  2,830   & 4 NU \\
     PSLV           & India     & 1.02\% &  3,250   &  9,540   & 2 Solid, 1 NU, 2 Solid \\
     Rockot         & Russia    & 2.00\% &  2,150   & 19,440   & 3 NU \\
     Soyuz FG       & Russia    & 2.56\% &  7,800   & 27,300   & 3 LO2K, 1 NU \\
     Soyuz 2.1a/b   & Russia    & 1.59\% &  4,850   & 16,500   & 3 LO2K, 1 NU \\
     Soyuz 2.1v     & Russia    & 1.91\% &  3,000   & 13,000   & 2 LO2K, 1 NU \\
     Vega           & France    & 1.47\% &  1,963   & 18,500   & 3 Solid, 1 NU \\
     Zenit          & Russia    & 1.31\% &  6,160   & 13,800   & 3 LO2K \\
     \hline
   \end{tabular}
   \captionof{table}{Launch Systems Used/Planned in 2017 \label{6.T01}}
\end{center}

\newpage
Fuel types used:
\begin{enumerate}
  \item Solid -- many rockets, developed technology.
  \item Liquid O$_2$/Kerosene -- many rockets, developed technology.
  \item Liquid O$_2$/Liquid H$_2$ -- many rockets, developed technology.
  \item N$_2$O$_4$/UDMH -- many rockets, developed technology.
  \item Liquid O$_2$/CH$_4$ -- SpaceX project; Black Arrow 2 project; New Glenn project; Soyuz 5 project; Vulcan project.
  \item Liquid O$_2$/Propylene -- Vector H project; Vector R project.
  \item H$_2$O$_2$/Kerosene -- Long March 6 final stage.
\end{enumerate}

\section{Inert Mass Ratio}

In the Table \ref{6.T02} below we list the \textbf{inert mass ratios} for the launch systems in use in 2018 \cite{Raport02}.  The first column is the launch vehicle name.  The first column is the launch vehicle name.  The second column is the nation launching the rocket.  The third column is the \textbf{inert mass ratios}, which is the quotient of combined empty mass of all stages to the liftoff mass.
The fourth column is the number and type of stages.  LO2K is liquid oxygen -- kerosene;  NU is N$_2$O$_4$/UDMH.  Rockets with solid boosters or liquid hydrogen propellant are excluded from the table.

\begin{center}
   \begin{tabular}{|l|l|l|r|r|l|l|l|l|}
     \hline
     Launch         & Nation    & Mass  & Stages        \\
     Vehicle        &           & Ratio &               \\
     \hline
     Angara 1.2     & Russia    & 7.1\% & 2 LO2K        \\
     Angara A3      & Russia    & 8.0\% & 2 LO2K, 1 NU  \\
     Angara A5      & Russia    & 7.3\% & 2 LO2K, 1 NU  \\
     Dnepr          & Russia    & 9.6\% & 3 NU          \\
     Electron       & USA       & 8.0\% & 2 LO2K        \\
     Long March 2C  & China     & 6.7\% & 2 NU          \\
     Long March 2F  & China     & 6.3\% & 3 NU          \\
     Long March 3B  & China     & 7.2\% & 3 NU          \\
     Long March 3C  & China     & 7.0\% & 3 NU          \\
     Proton M       & Russia    & 6.9\% & 4 NU          \\
     \hline
   \end{tabular}
   \captionof{table}{Inert Mass Ratio \label{6.T02}}
\end{center}

Isakowitz presented inert mass ratios for single stages used by 1991 \cite{LVS01}.  Out of 15 inert mass ratios, 1 are under 0.04, 6 are between 0.04 and 0.06, and 4 are between 0.06 and 0.08, and 4 are between 0.08 and 0.1.  Akin \cite{LVS02} has somewhat higher inert mass ratios for more recent LO2/Kerosene rocket stages.  Of the 8 mass ratios, 4 are between 0.04 and 0.06, 2 are between 0.06 and 0.08, and 2 are between 0.08 and 0.1.  The inert mass ratio for lighter rockets is generally a little higher than the one for heavier rockets.

Generally, for single use rocket stages with gross mass exceeding 50 tons, an inert mass ratio of 0.05 is sufficient.  We are designing a rocket stage which can withstand at least 4,000 fuelling-refuelling cycles.  First, the inert mass ratio exclusive of the heat shield for the first two stages should be about 0.12.  Second, the material used for the propellant tank must be fatigue-resistant.  Work with compressed natural gas cylinders has shown that some composite materials experience almost no fatigue \cite{CNG1}.  The effective inert mass ratio of the first two stages would include both the heat shield and the fuel used for deceleration and landing.

\section{Programs Used For Analysis of Reusable Launch System}
\begin{description}
  \item[FirstStageEquator.m] calculates the rocket's altitude, speed, and flight direction 10 seconds after burnout.  The user inputs MPDS Stage 1's propellant mass ratio, exhaust velocities at sea level and vacuum, and rocket thrust profile.

      Equations (\ref{3.01}) -- (\ref{3.13}) are used to calculate the ascent trajectory of MPDS Stage 1.  The Tsialkovski Equation (\ref{3.14}) is used to calculate the rocket velocity change.
      Equation (\ref{3.15}) is used to calculate the drag loss.
      Equation (\ref{3.16}) is used to calculate the gravity loss.
      Equation (\ref{3.17}) is used to calculate the vector loss.
  \item[SecondStageEquator.m] calculates the rocket's altitude, speed, and flight direction 10 seconds after burnout.  The user inputs MPDS Stage 2's propellant mass ratio, vacuum exhaust velocity, and rocket thrust profile.  The program obtains the rocket's altitude, speed, and flight direction calculated by \textbf{FirstStageEquator.m}.

      Similar to \textbf{FirstStageEquator.m}, Equations (\ref{3.01}) -- (\ref{3.13}) are used to calculate the ascent trajectory of MPDS Stage 2.  The Tsialkovski Equation (\ref{3.14}) is used to calculate the rocket velocity change.
      Equation (\ref{3.15}) is used to calculate the drag loss -- which is almost null for this stage.
      Equation (\ref{3.16}) is used to calculate the gravity loss -- which is significant but not null.
      Equation (\ref{3.17}) is used to calculate the vector loss -- which is also significant.
  \item[ThirdStageEquator.m] calculates the rocket's altitude, speed, and flight direction 10 seconds after burnout.  The user inputs MPTA Stage 1's propellant mass ratio, vacuum exhaust velocity, and rocket thrust profile.  The program obtains the rocket's altitude, speed, and flight direction calculated by \textbf{SecondStageEquator.m}.

      Equations used for previous two stages are used here.  Drag loss is null.  Gravity loss is very small.  Vector loss is significant.
  \item[FourthStageEquator.m] calculates  the rocket's altitude, speed, and flight direction 10 seconds after burnout.  The program also draws the orbit obtained by the payload. The user inputs MPTA Stage 2's propellant mass ratio, vacuum exhaust velocity, and rocket thrust profile.  The program obtains the rocket's altitude, speed, and flight direction calculated by \textbf{ThirdStageEquator.m}.

      Equations used for previous two stages are used here.  Drag loss is null.  Gravity loss is very small.  Vector loss is significant.  At the end of this stage, the payload settles into a Low Earth Orbit.
  \item[FirstStageReturn.m]  The program obtains MPDS Stage 1's altitude, speed, and flight direction obtained by \textbf{FirstStageEquator.m}.  Equations (\ref{3.01}) -- (\ref{3.13}) are used to calculate the descent trajectory of MPDS Stage 1.  At every time interval dt, the rocket's velocity vector, altitude, and horizontal coordinate is recorded.  The absolute values of acceleration are obtained from rocket velocity.  The density of surrounding air is obtained from altitude.

  \item[SecondStageReturn.m]  Similar to \textbf{FirstStageReturn.m}, the program obtains MPDS Stage 2's altitude, speed, and flight direction obtained by \textbf{SecondStageEquator.m}.  Equations (\ref{3.01}) -- (\ref{3.13}) are used to calculate the descent trajectory of MPDS Stage 2.  At every time interval dt, the rocket's velocity vector, altitude, and horizontal coordinate is recorded.  The absolute values of acceleration are obtained from rocket velocity.  The density of surrounding air is obtained from altitude.

      The difference with \textbf{FirstStageReturn.m} is that Stage 2 performs a retroburn and decreases mostly the vertical and slightly the horizontal components of its velocity.  The retroburn takes place at 70 $km$ -- 50 $km$ altitude as the rocket descends.
  \item[FlightAnalyze.m] analyzes the data obtained from \textbf{FirstStageReturn.m} or \textbf{SecondStageReturn.m}.  The data consists of the rocket's vector velocity, ambient air density, and ambient air temperature at every point $dt$ during return.  The program plots the rocket's speed and deceleration.  The program calculates the heat flux on the cap and rocket cylinder.  \textbf{FlightAnalyze.m} keeps track of the heat shield temperature.  The program calculates the total thermal energy absorbed by the heat shield.

      First, \textbf{FlightAnalyze.m} calculates laminar flow heat flux.  The heat flux at the stagnation point is calculated by (\ref{5.09}).

      Second, \textbf{FlightAnalyze.m} calculates turbulent flow heat flux.  The heat flux on the cylinder sides is calculated by (\ref{5.44}).  The thermal energy absorbed by the caps is calculated by (\ref{5.53}).
\end{description}

\section{Abbreviations}
\subsection{General}
\begin{tabular}{lcl}
\textbf{CCT}     & -- & average temperature in the combustion chamber\\
\textbf{CCP}     & -- & pressure in the combustion chamber\\
\textbf{EAR}     & -- & area ratio of the nozzle exit to the nozzle throat (expansion area ratio)\\
\textbf{LEO}     & -- & low Earth orbit\\
\textbf{MPDS}    & -- & Midpoint Delivery System -- delivers the Midpoint to Orbit Delivery System \\
& & (MPTO) and the payload to a midpoint in orbital ascent\\
\textbf{MPTO}    & -- & Midpoint to Orbit Delivery System -- delivers the payload from a midpoint\\
& & to orbit\\
\textbf{NAFMF}   & -- & the mass fraction of everything except the fuel mass used in the ascent\\
\textbf{NED}     & -- & rocket nozzle exit diameter\\
\textbf{ODS}     & -- & Orbital Delivery System \\
\textbf{OFR}     & -- & the mass ratio of oxidizer to fuel\\
\textbf{RPA}     & -- & rocket propulsion analysis\\
\textbf{SSTO}    & -- & single stage to orbit\\
\textbf{VTVL}    & -- & vertical takeoff, vertical landing
\end{tabular}

\subsection{Chemicals}
\begin{tabular}{lcl}
\textbf{AcetAm40} & -- & a solution of 60\% Ammonia and 40\% Acetylene by mass\\
\textbf{AN}       & -- & ammonium nitrate\\
\textbf{CNG}      & -- & compressed natural gas\\
\textbf{EEC}      & –- & the mixture of 61\% monoethanolamine, 30\% ethanol, and 9\% hydrated \\ & & copper nitrate\\
\textbf{ETA}      & –- & the mixture of ethanolamine and 10\% CuCl$_2$\\
\textbf{ETAFA}    & –- & the mixture of 47.5\% Ethanolamine, 47.5\% Furfuryl Alcohol, and \\
& & 5.0\% CuCl$_2$\\
\textbf{EthOx40}  & -- & a solution of 40\% ethyl oxide and 60\% water\\
\textbf{HP95}     & -- & 95\% hydrogen peroxide\\
\textbf{LO2}      & -- & liquid oxygen\\
\textbf{RP-1}     & -- & rocket propellant (kerosene)\\
\end{tabular}
\section{List of Symbols}
\subsection{Greek}
\begin{flushleft}
\begin{tabular}{lcl}
  $\gamma$  & -- & adiabatic constant of air \\
  $\eta_{_{\text{Subscript}}}$  & -- & efficiency \\
  $\theta$  & -- & angle of inclination with respect to hypersonic flow in Section 4 \\
  $\mu$  & -- & air viscosity \\
  $\mu^*$  & -- & air viscosity at the vehicle wall boundary \\
  $\mu_a$  & -- & ambient air viscosity \\
  $\rho$  & -- & air density; in Subsection 3.2 used as the density of propellant;\\
  $\sigma_B$  & -- & Stefan -- Boltzmann constant \\
  $\sigma_{_{\text{HSS}}}$ & -- & areal density of the heat shield given in (\ref{5.57}) \\
\end{tabular}
\end{flushleft}

\subsection{Latin}
\begin{flushleft}
\begin{longtable}{lcl}
  $\mathbf{a}_{_r}(t)$ & -- & magnitude and direction of the acceleration due to rocket engines experienced\\
  & &  by the rocket at time $t$ \\
  $A$ & -- & area \\
  $C_{_d}$ & -- & drag coefficient \\
  $C_f(x)$ & -- & skin friction coefficient at distance $x$ from the leading edge \\
  $e_{_w}$ & -- & wall emissivity \\
  $\mathcal{E}$ & -- & total kinetic and potential energy per unit rocket (projectile) mass \\
  $E_{_{\text{Subscript}}}^{^{\text{Superscript}}}$ & -- & total thermal energy absorbed (Chapter 4) \\
  $\mathbf{F}_{_{\text{Subscript}}}$ & -- & general force \\
  $\mathbf{F}_{_d}$ & -- & air drag force \\
  $\mathbf{F}_{_t}$ & -- & rocket thrust force \\
  $\textbf{g}$ & -- & directed acceleration of Earth's gravity \\
  $\hat{g}$ & -- & direction of acceleration of Earth's gravity \\
  $g_{_0}$ & -- & scalar acceleration of Earth's gravity, 9.81 $m/s^2$ \\
  $h$ & -- & altitude \\
  $h_{_{\text{Subscript}}}$ & -- & may be used for altitude or enthalpy or energy per unit mass \\
  $H^*$ & -- & air enthalpy at the vehicle wall boundary \\
  $H_{_a}$ & -- & ambient air enthalpy \\
  $H_{_{aw}}$ & -- & air enthalpy at adiabatic wall temperature \\
  $H_{_w}$ & -- & air enthalpy at wall temperature \\
  $I$ & -- & impulse \\
  $I_{_{sp}}$ & -- & specific impulse \\
  $L/D$ & -- & rocket length to diameter ration (in most other works, it is lift to drag ratio) \\
  $M$ & -- & rocket or hypersonic vehicle mass \\
  $\mathcal{M}$ & -- & Mach number \\
  $\mathcal{M}_{_w}$ & -- & the Mach number of air close to the wall relative to the rocket \\
  $P$ & -- & pressure \\
  $P_{_{CC}}$ & -- & combustion chamber pressure \\
  $Pr$ & -- & Prandtl number \\
  $Q_{_{\text{Subscript}}}^{^{\text{Superscript}}}$ & -- & heat load \\
  $\dot{Q}_{_{\text{Subscript}}}^{^{\text{Superscript}}}$ & -- & heat flux \\
  $\dot{Q}_{_{K}}$ & -- & heating constant given in (\ref{5.04}) -- (\ref{5.06}) \\
  $r_{_f}$ & -- & recovery factor \\
  $R_{_E}$ & -- & Earth's radius \\
  $Re$ & -- & Reynolds number \\
  $S_{_f}$ & -- & the safety factor or the ratio of minimal burst pressure to designed tank pressure \\
  $\mathcal{S}$ & -- & specific yield strength \\
  $St$ & -- & Stanton number \\
  $T^*$ & -- & air temperature at the vehicle wall boundary \\
  $T_{_a}$ & -- & ambient air temperature \\
  $T_{_r}$ & -- & recovery temperature \\
  $T_{_w}$ & -- & wall temperature \\
  $v$ & -- & velocity \\
  $\bar{v}_{_e}$ & -- & exhaust velocity \\
  $v_{_e}$ & -- & average exhaust velocity \\
  $v_{_{e0}}$ & -- & ideal exhaust velocity calculated by Rocket Propulsion Analysis software \\
  $v_{_E}(\theta)$ & -- & velocity of Earth's rotation \\
  $v_{_g}$ & -- & rocket velocity gain \\
  $v_{_s}$ & -- & speed of sound \\
  $\triangle v$ & -- & velocity change due to rocket engine \\
  $\triangle v_{_d}$ & -- & air drag velocity loss \\
  $\triangle v_{_g}$ & -- & gravity velocity loss \\
  $\triangle v_{_v}$ & -- & vector velocity loss \\
  $V$ & -- & volume \\
  $y$ & -- & dimensionless non-angular coordinate on the cylindrically symmetric\\
   & & rocket surface, $ry$ is the distance from the stagnation point in meters \\
\end{longtable}
\end{flushleft}

\end{document}